\documentclass[twocolumn,english,aps,prl,onecolumn,superscriptaddress]{revtex4}
\usepackage[T1]{fontenc}
\usepackage[latin9]{inputenc}
\usepackage{amsmath}
\usepackage{amssymb}
\usepackage{stackrel}
\usepackage{graphicx}
\usepackage{babel}
\usepackage[colorlinks,urlcolor=blue,linkcolor=blue,citecolor=blue]{hyperref}
\usepackage{docmute}
\makeatletter
\@ifundefined{textcolor}{}
{%
 \definecolor{BLACK}{gray}{0}
 \definecolor{WHITE}{gray}{1}
 \definecolor{RED}{rgb}{1,0,0}
 \definecolor{GREEN}{rgb}{0,1,0}
 \definecolor{BLUE}{rgb}{0,0,1}
 \definecolor{CYAN}{cmyk}{1,0,0,0}
 \definecolor{MAGENTA}{cmyk}{0,1,0,0}
 \definecolor{YELLOW}{cmyk}{0,0,1,0}
}

\makeatother

\begin{document}

\title{Broad-Band Negative Refraction via Simultaneous Multi-Electron Transitions}

\author{Jing-Jing Cheng}

\address{Department of Physics, Applied Optics Beijing Area Major Laboratory,
Beijing Normal University, Beijing 100875, China}

\author{Ying-Qi Chu}

\address{Department of Physics, Applied Optics Beijing Area Major Laboratory,
Beijing Normal University, Beijing 100875, China}

\author{Tao Liu}

\address{Theoretical Quantum Physics Laboratory, RIKEN Cluster for Pioneering
Research, Wako-shi, Saitama 351-0198, Japan}

\author{Jie-Xing Zhao}

\address{Department of Physics, Applied Optics Beijing Area Major Laboratory,
Beijing Normal University, Beijing 100875, China}

\author{Fu-Guo Deng}

\address{Department of Physics, Applied Optics Beijing Area Major Laboratory,
Beijing Normal University, Beijing 100875, China}
\address{NAAM-Research Group, Department of Mathematics, Faculty of Science,
King Abdulaziz University, Jeddah 21589, Saudi Arabia}

\author{Qing Ai}
\email{aiqing@bnu.edu.cn}

\address{Theoretical Quantum Physics Laboratory, RIKEN Cluster for Pioneering
Research, Wako-shi, Saitama 351-0198, Japan}

\address{Department of Physics, Applied Optics Beijing Area Major Laboratory,
Beijing Normal University, Beijing 100875, China}

\author{Franco Nori}
\email{fnori@riken.jp}

\address{Theoretical Quantum Physics Laboratory, RIKEN Cluster for Pioneering
Research, Wako-shi, Saitama 351-0198, Japan}

\address{Department of Physics, The University of Michigan, Ann Arbor, Michigan
48109-1040, USA}

\date{\today}
\begin{abstract}
We analyze different factors which influence the negative refraction
in solids and multi-atom molecules. We find that
this negative refraction is significantly influenced
by simultaneous multi-electron transitions
with the same transition frequency and dipole redistribution
over different eigenstates. We show that these simultaneous multi-electron transitions
and enhanced transition dipole \textit{broaden} the bandwidth of the
negative refraction by at least \textit{one order of magnitude}.
This work provides additional connection between metamaterials and M\"{o}bius strips.
\end{abstract}
\maketitle

\textit{Introduction}.\textendash{}\textendash{}
Metamaterials with negative refraction \cite{Veselago1968,Pendry2000,Smith2000,Bliokh2008,Khorasaninejad2017,Minovich2015,Zhao2016}
have attracted broad interest because of its potential applications,
including perfect lenses \cite{Pendry2000},
fingerprint identification in forensic science \cite{Shen2016},
simulating condensed matter phenomena and reversed
Doppler effect \cite{Bliokh2013,Kats2007},
controlling light's polarization \cite{Jiang2014,Fan2015},
and electromagnetic cloaking \cite{Leonhard2006,Pendry2006}.
In order to realize metamaterials,
a number of routes have been proposed, including (molecular)
split-ring resonators \cite{Smith2000,Xiong2013,Shen2014},
chiral routes \cite{Pendry2004,Xiong2010},
hyperbolic dispersion \cite{Fisher1969,Smith2003,Rakhmanov2010,Ai2018},
dark-state mechanism \cite{Kastel2007-1,Kastel2007-2,Qin2013,Wang2018},
and topological routes \cite{Xiong2009,Chang2010,Krishnamoorthy2012,Fang2016}.
However, none of these can effectively
overcome the difficulty of realizing broad-band negative refraction.

In multi-atom molecules and solids, there are generally more than
one electron which can respond to applied electromagnetic fields.
In previous investigations \cite{Shen2014,Shen2016,Fang2016,Oktel2004,Kastel2007-1,Kastel2007-2,Chen2005,Orth2013,Thommen2006,Ceulemans1998,Zagoskin2015}, only transitions for a \textit{single} electron were considered. Because each atom
contributes a $\pi$-electron to the molecule, there are $N$ $\pi$-electrons
in a molecule with $N$ atoms. Thus, there would be $N$ possible
transitions for the $\pi$-electrons in the molecule. Since two or more transitions can effectively overlap with each other, it would be reasonable to expect a broadened bandwidth of the negative refraction by simultaneous multi-electron transitions. On the other hand, M\"{o}bius molecules with nontrivial topology have been successfully synthesized \cite{Heilbronner1964,Walba1993,Ajami2003,Yoneda2014} and proposed to realize, e.g., metamaterials \cite{Chang2010,Fang2016,Poddar2014}, quantum devices \cite{Balzani2008,Yamashiroa2004,Zhao2009}, dual-mode resonators and bandpass filters \cite{Pond2000}, topological insulators \cite{Guo2009}, molecular knots and engines \cite{Lukin2005}, and artificial light harvesting \cite{Xu2018,Lambert2013}. As a concrete example for demonstrating the above principle, here we consider multi-electron transitions in a double-ring M\"{o}bius molecule \cite{Zhao2009,Fang2016}
with $2N$ $\pi$-electrons. Every energy
level could be filled with two electrons of different spins, due to Pauli's exclusion
principle. Because the lower energy level has a higher probability
to be filled with electrons, half of the $N$ energy levels with lower energy can
be filled with electrons. Furthermore, electrons
can transit from an occupied energy level to an unoccupied energy level.
Because multi-electrons can be involved in the transitions of nearby frequencies,
the bandwidth of the negative refraction would be \textit{significantly broadened} compared to previous findings \cite{Kastel2007-1,Kastel2007-2,Fang2016,Thommen2006,Oktel2004},
as illustrated in Fig.~\ref{fig:scheme}.

Generally, it is difficult to fabricate metamaterials,
as most of these require producing a huge number of split-ring resonators.
Also, the size of a resonator in classical metamaterials is
of the order of the wavelength of the electromagnetic field \cite{Landau1995,Jackson1999}.
However, since in a molecular medium the electromagnetic response
results from quantum transitions, the size of a molecule for quantum metamaterials
can be intrinsically smaller by 2--3 orders of magnitude than the size of a resonator for classical metamaterials. Furthermore, metamaterials consisting of molecules are crystal of such molecules and thus can be easily fabricated by crystallization.

\textit{Simultaneous Multi-Electron Transitions}.\textendash{}\textendash{}The electric displacement field $\vec{D}$ and magnetic induction $\vec{B}$ can be written as \cite{Landau1995,Jackson1999}
\begin{eqnarray}
\vec{D} & = & \varepsilon_{0}\vec{E}+\vec{P}=\varepsilon_{0}\overleftrightarrow{\varepsilon_{r}}\vec{E},
\label{eq:D} \\
\vec{B} & = & \mu_{0}(\vec{H}+\vec{M})=\mu_{0}\overleftrightarrow{\mu_{r}}\vec{H},\label{eq:B}
\end{eqnarray}
where $\vec{E}$ is the applied electric field, $\vec{P}$ is the polarization,
$\varepsilon_{0}$ and $\varepsilon_{0}\overleftrightarrow{\varepsilon_{r}}$ are, respectively, the permittivity of the vacuum and medium,
$\vec{H}$ is the applied magnetic field,
$\vec{M}$ is the magnetization, $\mu_{0}$ and $\mu_{0}\overleftrightarrow{\mu_{r}}$ are, respectively, the permeability of the vacuum and medium.
Under the dipole approximation, according to linear response theory \cite{Kubo1985}, the relative permittivity and permeability are given by \cite{Shen2014,Shen2016,Fang2016,Ai2018,SuppMat}
\begin{eqnarray}
\overleftrightarrow{\varepsilon_{r}}&=&1-\sum_{i\neq f}\frac{\vec{d}_{if}
\vec{d}_{fi}}{n_{i}\hbar\varepsilon_{0}v_{0}}\;\textrm{Re}\!\left(\frac{1}
{\omega-\Delta_{fi}+i\gamma}\right),\label{eq:Permit}\\
\overleftrightarrow{\mu_{r}}&=&1-\sum_{i\neq f}\frac{\mu_{0}\vec{m}_{if}
\vec{m}_{fi}}{n_{i}\hbar v_{0}}\;\textrm{Re}\!\left(\frac{1}
{\omega-\Delta_{fi}+i\gamma}\right),\label{eq:Permea}
\end{eqnarray}
where $\vec{d}_{if}$ ($\vec{m}_{if}$) is the transition electric (magnetic) dipole between the initial state $\vert i\rangle$ and the final state $\vert f\rangle$ with level spacing $\hbar\Delta_{fi}$, $n_i^{-1}$ is the number of electrons occupying the initial state,
$\omega$ is the frequency of the applied electromagnetic field,
$\hbar$ is the Planck constant, and $v_0$ is the volume of the molecule. In obtaining Eqs.~(\ref{eq:Permit},\ref{eq:Permea}), we assume that all molecules are identical, and every molecule responds equally to the applied electromagnetic field.
Due to the response of the molecules,
the actual electromagnetic field inside the medium is different from that in the vacuum.
However, it can be proven by Lorentzian local field theory that both the center and bandwidth of the negative refraction will not be modified significantly \cite{Kastel2007-1,Kastel2007-2,Fang2016,Ai2018}. Furthermore, the dipole-dipole interaction between molecules can be omitted as long as it is smaller than either
the decoherence rate of the molecular excited states \cite{Jang2018,Novoderezhkin2010} or the static disorder \cite{Cheng2006}.

As a specific case, we consider a M\"{o}bius molecule with double rings consisting of $2N$
carbon atoms \cite{Ajami2003}, as shown in Fig.~\ref{fig:scheme}.
The $j$th atomic position of ring a (b) is $\vec{R}_{j+}$ ($\vec{R}_{j-}$) with
\begin{eqnarray}
\vec{R}_{j\pm} & = & \left(R\pm W\sin\!\frac{\varphi_{j}}{2}\right)\cos\varphi_{j}\hat{e}_{x}\pm W\cos\frac{\varphi_{j}}{2}\hat{e}_{z}\nonumber \\
 &  & +\left(R\pm W\sin\frac{\varphi_{j}}{2}\right)\sin\varphi_{j}\hat{e}_{y},
\end{eqnarray}
where the radius and width of the M\"{o}bius molecule are, respectively, $R$ and
$2W$, with $W$ the radius of the carbon atom, $\varphi_{j}=j\delta$, and $\delta=2\pi/N$.
According to the H\"{u}ckel molecular orbital theory \cite{Salem1972}, the Hamiltonian of the M\"{o}bius molecule for the single electron is \cite{Zhao2009,Fang2016}
\begin{eqnarray}
H & = & \sum_{j=0}^{N-1}\left[A_{j}^{\dagger}MA_{j}-\xi\left(A_{j}^{\dagger}A_{j+1}+\mathrm{h.c.}\right)\right],
\end{eqnarray}
where $A_{j}=(a_{j},b_{j})^T$ with $a_{j}$ ($b_{j}$) the annihilation operator at the $j$th site of ring a (b),
\begin{align}
M & =\left[\begin{array}{cc}
\epsilon & -V\\
-V & -\epsilon
\end{array}\right],
\end{align}
$\epsilon$ is the on-site energy difference between
the two rings, and $V$ ($\xi$) is the inter-ring (intra-ring)
resonant integral. Because of the M\"{o}bius boundary condition,
the last atom of each ring is linked to the first atom of the other
ring, i.e., $a_{0}=b_{N}$, and $b_{0}=a_{N}$.

\begin{figure}
\includegraphics[bb=96 239 516 554,width=8cm]{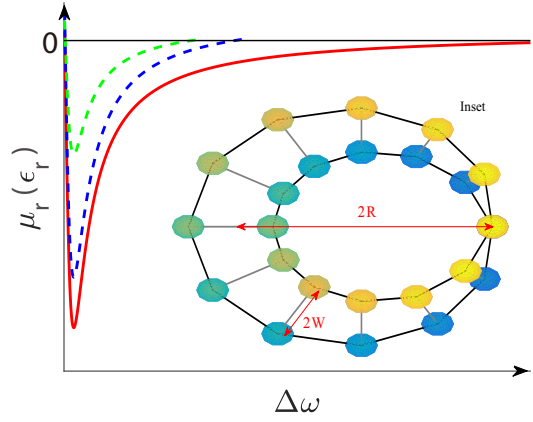}
\caption{Schematic illustration of broad-band negative refraction induced by simultaneous  multi-electron transitions. Dashed curves are for single-electron transitions,
while the solid curve provides the overall simultaneous contribution.
Inset: Top view of a M\"{o}bius molecule with $N=12$ carbon atoms in each ring.
The adjacent atoms are linked with covalent bonds.
}
\label{fig:scheme}
\end{figure}

The M\"{o}bius boundary condition can be canceled by a local unitary transformation, i.e.,
$B_{j} \equiv (c_{j\uparrow},c_{j\downarrow})^T=U_{j}A_{j}$, $B_{N}=B_{0}$,
where $c_{j\sigma}$ is the annihilation operator of an electron at
$j$th atomic site with $\sigma$ pseudo-spin label,
\begin{align}
U_{j} & =\frac{1}{\sqrt{2}}\left[\begin{array}{cc}
e^{-i\varphi_{j}/2} & -e^{-i\varphi_{j}/2}\\
1 & 1
\end{array}\right].
\end{align}
Therefore, the Hamiltonian with periodic boundary condition can be rewritten as
\begin{equation}
H=\sum_{j=0}^{N-1}\left[B_{j}^{\dagger}V\sigma_{z}B_{j}-\xi\left(B_{j}^{\dagger}QB_{j+1}+\textrm{h.c.}\right)\right],
\end{equation}
where
\begin{align}
Q & =\left[\begin{array}{cc}
e^{i\delta/2} & 0\\
0 & 1
\end{array}\right].
\end{align}
By using the Fourier transform,
$B_{j}=\textstyle\sum_{k}e^{-ikj}C_{k}$,
the Hamiltonian of the M\"{o}bius molecule is diagonalized as
\begin{equation}
H=C_{k}^{\dagger}E_{k}C_{k},
\end{equation}
where $C_{k} =(c_{k\uparrow},c_{k\downarrow})^T$ with
\begin{align}
c_{k\uparrow}  & =\frac{1}{\sqrt{2N}}\sum_{j=0}^{N-1}e^{-i(k-\delta/2)j}
\left(a_{j}^{\dagger}-b_{j}^{\dagger}\right) ,\\
c_{k\downarrow}  & =\frac{1}{\sqrt{2N}}\sum_{j=0}^{N-1}e^{-ikj}
\left(a_{j}^{\dagger}+b_{j}^{\dagger}\right) ,\\
E_{k} & =\left[\begin{array}{cc}
E_{k\uparrow} & 0\\
0 & E_{k\downarrow}
\end{array}\right]\label{eq:Ek}
\end{align}
with $E_{k\uparrow}=V-2\xi\cos\left(k-\frac{\delta}{2}\right)$
and $E_{k\downarrow}=-V-2\xi\cos k$ the eigenenergies of the upper and lower bands, respectively.

\textit{Selection Rules for Transitions}.\textendash{}\textendash{}In Eqs.~(\ref{eq:Permit},\ref{eq:Permea}),
we apply linear response theory to rewrite
the relative permittivity and permeability in terms of the transition
matrix elements of the electric and magnetic dipoles, i.e.,
$\vec{d}{}_{if}$ and $\vec{m}{}_{if}$. According to
Eqs.~(\ref{eq:Permit},\ref{eq:Permea}), we would expect negative
permittivity and permeability close to the transition frequencies $\Delta_{fi}=(E_f-E_i)/\hbar$.
The selection rules of transitions are explicitly provided
by the matrix elements of the electric and magnetic dipoles \cite{SuppMat}.

Using the dipole approximation, the interaction Hamiltonian with the electric
field is written as \cite{Landau1995,Jackson1999}
\begin{equation}
H_{E}^{\prime}=-\vec{d}\cdot\vec{E}_{0}\textrm{cos}\omega t,
\end{equation}
where the molecule is subject to an electric field with
amplitude $\vec{E}_{0}=(E_{0}^{x},E_{0}^{y},E_{0}^{z})$ and frequency $\omega$.
The electric-dipole-induced transition is allowed if the corresponding
matrix element is nonzero. By using the relation $\left\langle \phi_{js}\right|\vec{r}\left|\phi_{j's'}\right\rangle =\vec{R}_{js}\delta_{jj^{\prime}}\delta_{ss^{\prime}}$ \cite{Salem1972} and the rotating-wave approximation \cite{Ai2010,Wang2017}, the selection rules for the
electric-dipole-induced transitions are summarized as
\begin{subequations}\label{eq:ElecSelRul}
\begin{eqnarray}
\left|k,\downarrow\right\rangle  & \overset{x,y,z}{\rightleftharpoons} & \left|k,\uparrow\right\rangle ,\\
\left|k,\downarrow\right\rangle  & \overset{x,y}{\rightleftharpoons} & \left|k+2\delta,\uparrow\right\rangle ,\\
\left|k,\downarrow\right\rangle  & \overset{x,y,z}{\rightleftharpoons} & \left|k+\delta,\uparrow\right\rangle ,\\
\left|k,\downarrow\right\rangle  & \overset{x,y}{\rightleftharpoons} & \left|k-\delta,\uparrow\right\rangle ,
\end{eqnarray}
\end{subequations}
where the superscripts over the arrows indicate
the polarization of the electric field.
Under the dipole approximation, the interaction Hamiltonian with the magnetic field
is written as \cite{Landau1995,Jackson1999}
\begin{equation}
H_{B}^{\prime}=-\vec{m}\cdot\vec{B}_{0}\textrm{cos}\omega t,
\end{equation}
where $\vec{B}_{0}=(B_{0}^{x},B_{0}^{y},B_{0}^{z})$.
A straightforward calculation shows \cite{SuppMat} that the selection rules for
the magnetic-dipole-induced transitions are the same as
Eq.~(\ref{eq:ElecSelRul}).
The transitions allowed by both electric and magnetic dipole couplings
are depicted in Fig.~\ref{fig:SelectionRule},
in combination with initial and final conditions.
Because a transition can take place from the highest occupied molecular orbital (HOMO),
denoted by solid symbols,
to the lowest unoccupied molecular orbital (LUMO),
denoted by empty symbols.

\begin{figure}
\includegraphics[bb=20 0 385 300,width=8.5cm]{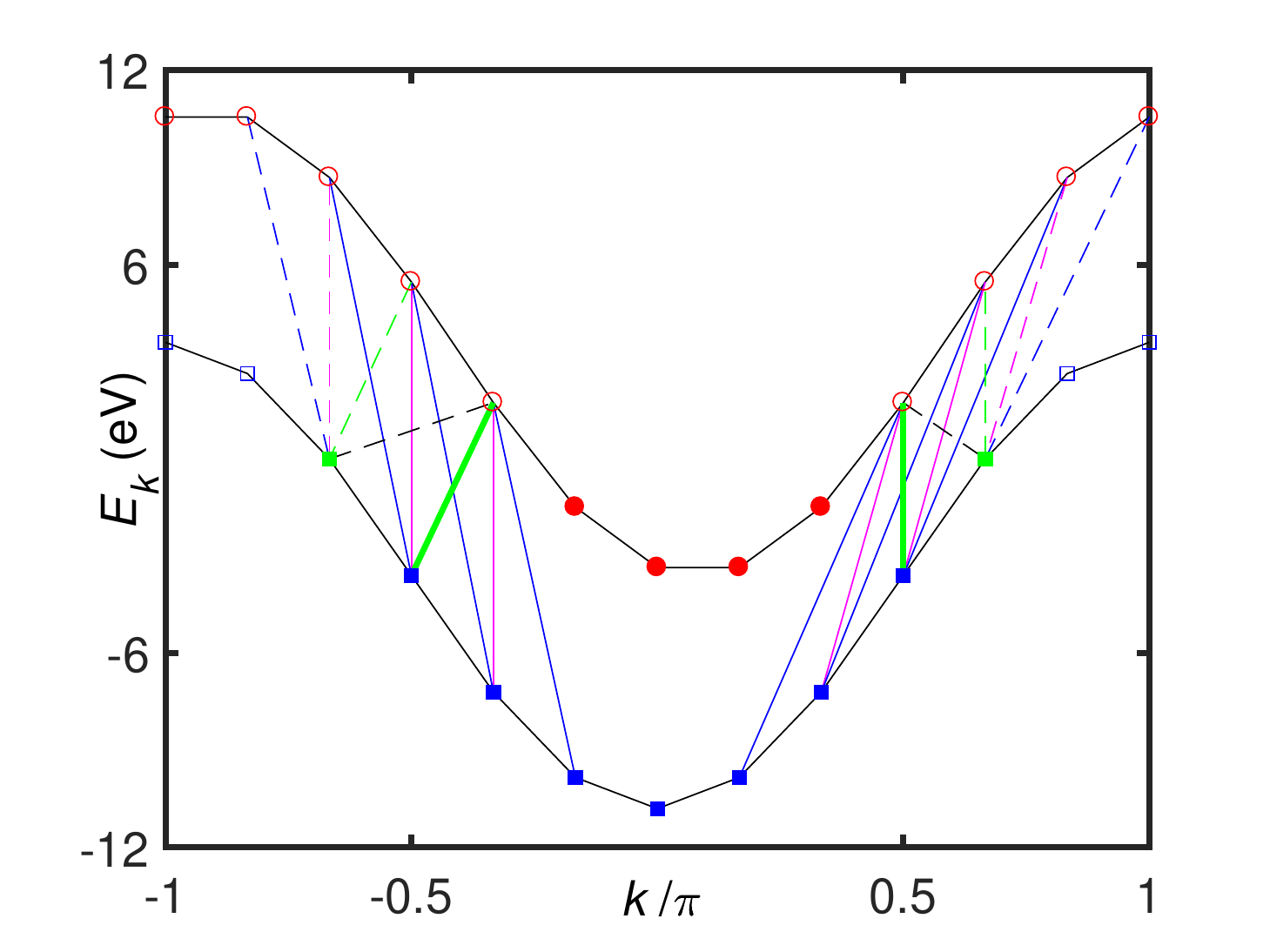}
\caption{ Selection rules for transitions in a M\"{o}bius
molecule. The curve marked with circles (squares) represents the upper (lower) energy
band. The red circles and blue squares are the levels occupied
by two $\pi$ electrons with spin up and down, while the empty symbols
are unoccupied. The green squares are occupied by one $\pi$
electron. The curves connecting two bands indicate the selection rules for transitions. Hereafter, we choose the following parameters: $\gamma^{-1}=4$~ns~\cite{Fang2016}, $V=\xi=3.6$~eV \cite{Greenwood1972}, $W=0.077$~nm \cite{Yoneda2014}, and $R=NW/\pi$.}
\label{fig:SelectionRule}
\end{figure}

\textit{Broad-Band Negative Refraction}.\textendash{}\textendash{}In order to
investigate the effects of simultaneous multi-electron transitions on the negative refraction,
we explore the relative permittivity and permeability for different detunings $\Delta\omega=\omega-\Delta_{fi}$.
In Eq.~(\ref{eq:Ek}), the lower-band $E_{k,\downarrow}$ is symmetric with respect to $k=0$, while the upper-band $E_{k,\uparrow}$ is symmetric with respect to $k=\delta/2$.
The following four pairs of transitions possess the same transition frequencies, respectively: $\vert k,\downarrow\rangle\leftrightarrows\vert k,\uparrow\rangle$ and
$\vert -k,\downarrow\rangle\leftrightarrows\vert -k+\delta,\uparrow\rangle$,
denoted by the green curves in Fig~\ref{fig:SelectionRule};
$\vert k,\downarrow\rangle\leftrightarrows\vert k+\delta,\uparrow\rangle$ and
$\vert-k,\downarrow\rangle\leftrightarrows\vert-k,\uparrow\rangle$,
denoted by the magenta curves;
$\vert k,\downarrow\rangle\leftrightarrows\vert k-\delta,\uparrow\rangle$ and
$\vert -k,\downarrow\rangle\leftrightarrows\vert -k+2\delta,\uparrow\rangle$,
denoted by the black curves;
$\vert k,\downarrow\rangle\leftrightarrows\vert k+2\delta,\uparrow\rangle$ and
$\vert -k,\downarrow\rangle\leftrightarrows\vert -k-\delta,\uparrow\rangle$,
denoted by the blue curves, which also fulfill the selection rules (\ref{eq:ElecSelRul}) for the transitions in M\"{o}bius molecules.

\begin{figure}
\includegraphics[bb=10 0 420 235,width=8.5cm]{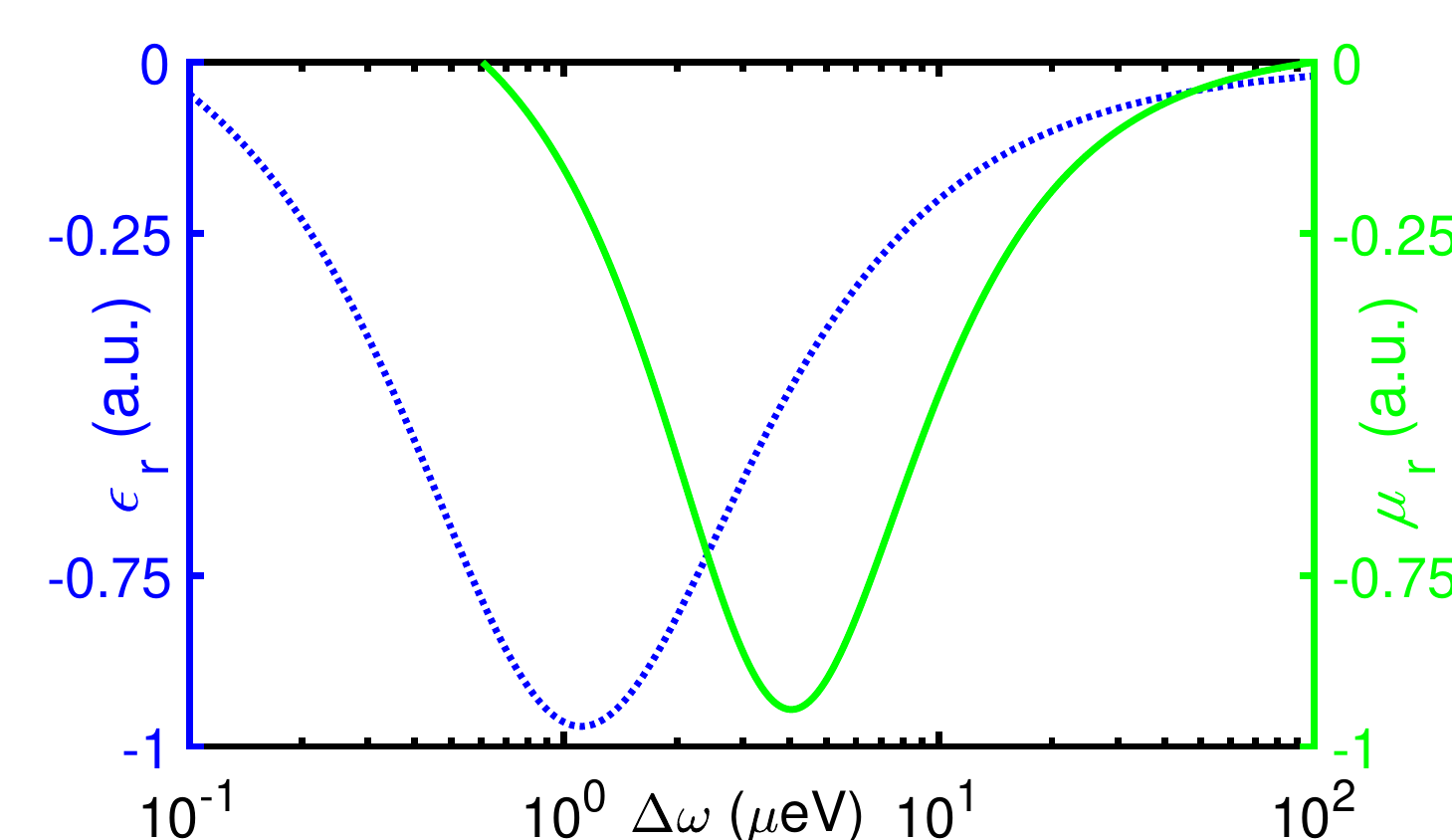}
\caption{ Relative permittivity $\varepsilon_r$ (blue dotted curve) and permeability $\mu_r$ (green curve) of a M\"{o}bius molecule versus the detuning $\Delta\omega$. Here we only show the lowest eigenvalue.
}
\label{fig:Widest}
\end{figure}

In Fig.~\ref{fig:Widest}, we show the widest bandwidth for the negative refraction which corresponds to the transitions labeled by the thick green curves in Fig.~\ref{fig:SelectionRule}. The bandwidth of the negative refraction is 80~$\mu$eV,
which is about \textit{20 times} that of Ref.~\cite{Fang2016}.
To explore the underlying broadening mechanism,
we investigate the relevant contributions from all simultaneous transitions.
Generally, the magnetic response is much smaller than the electric response.
Therefore, the bandwidth of the negative refraction is mainly determined by bandwidth of the negative permeability. In Fig.~\ref{fig:Contribution}, we investigate the effect of simultaneous transitions on the negative refraction by the relative permeability.
At the transition frequency, there are two degenerate transitions,
denoted by the thick green curves in Fig.~\ref{fig:SelectionRule}.
One is from the initial state $\left\vert-\frac{\pi}{2},\downarrow\right\rangle$
to the final state $\left\vert-\frac{\pi}{3},\uparrow\right\rangle$,
denoted by the green circles in Fig.~\ref{fig:Contribution}. The other transition is from
$\left\vert\frac{\pi}{2},\downarrow\right\rangle$
to $\left\vert\frac{\pi}{2},\uparrow\right\rangle$, denoted by the black curve.
The summation of the two transitions, denoted by the red dotted curve,
is nearly the same as that of all contributions.

The bandwidth of the negative refraction is determined by the zeros of the lowest eigenvalue of $\overleftrightarrow{\mu_{r}}$ \cite{SuppMat}, i.e.,
\begin{align}
\mu_{r}^1=1-\sum_{k\in\textrm{HOMO}}\;\sum_{k'\in\textrm{LUMO}}2\alpha_{k,k'}^2\eta'_{k,k'}=0,
\end{align}
where $\eta'_{k,k'}$ is of the same form for different transitions
but with a different central frequency.
Among four possible $\alpha_{k,k'}$'s,
the maximum of $\alpha_{k,k}$ and $\alpha_{k,k+\delta}$ are generally larger than those of $\alpha_{k,k-\delta}$ and $\alpha_{k,k+2\delta}$ \cite{SuppMat}. Assuming that $\varepsilon=V$, we have
\begin{eqnarray}
\alpha_{\frac{\pi}{2},\frac{\pi}{2}}
&\simeq&\frac{RV}{\hbar c}\left(1+2\sin\frac{3\delta}{4}\right)\simeq\frac{7RV}{4\hbar c},\\
\alpha_{-\frac{\pi}{2},-\frac{\pi}{3}}
&\simeq&\frac{RV}{\hbar c}\left(1+2\sin\frac{3\delta}{4}\right)\simeq\frac{7RV}{4\hbar c}.
\end{eqnarray}
In Ref.~\cite{Fang2016}, negative refraction was considered for the single transition
$\vert0,\downarrow\rangle\rightleftarrows\vert0,\uparrow\rangle$ with
\begin{eqnarray}
\alpha_{0,0}\simeq\frac{RV}{\hbar c}\left(1-2\sin\frac{\delta}{2}\sin\frac{3\delta}{4}\right)\simeq\frac{13RV}{16\hbar c}.
\end{eqnarray}
Since
$\alpha_{\frac{\pi}{2},\frac{\pi}{2}}^2=
\alpha_{-\frac{\pi}{2},-\frac{\pi}{3}}^2\simeq5\alpha_{0,0}^2$,
the transition dipoles have been enlarged by a factor of $\sqrt{5}$.
Furthermore, there are four simultaneous transitions with the same transition frequency, i.e., $\left\vert\frac{\pi}{2},\downarrow\right\rangle\rightleftarrows\left\vert\frac{\pi}{2},\uparrow\right\rangle$ and
$\left\vert-\frac{\pi}{2},\downarrow\right\rangle\rightleftarrows\left\vert-\frac{\pi}{3},\uparrow\right\rangle$ with two different electronic spins,
the bandwidth of the negative refraction is about 20 times of our previous observation in Ref.~\cite{Fang2016}. Compared to other discoveries, our proposed bandwidth is about:
$\mathit{10^5}$ \textit{times} that of Refs.~\cite{Kastel2007-1,Kastel2007-2}, $\mathit{4\times10^5}$ \textit{times} that of Ref.~\cite{Oktel2004}, and \textit{121 times} that of Ref.~\cite{Thommen2006}.

\begin{figure}
\includegraphics[bb=10 0 420 240,width=9cm]{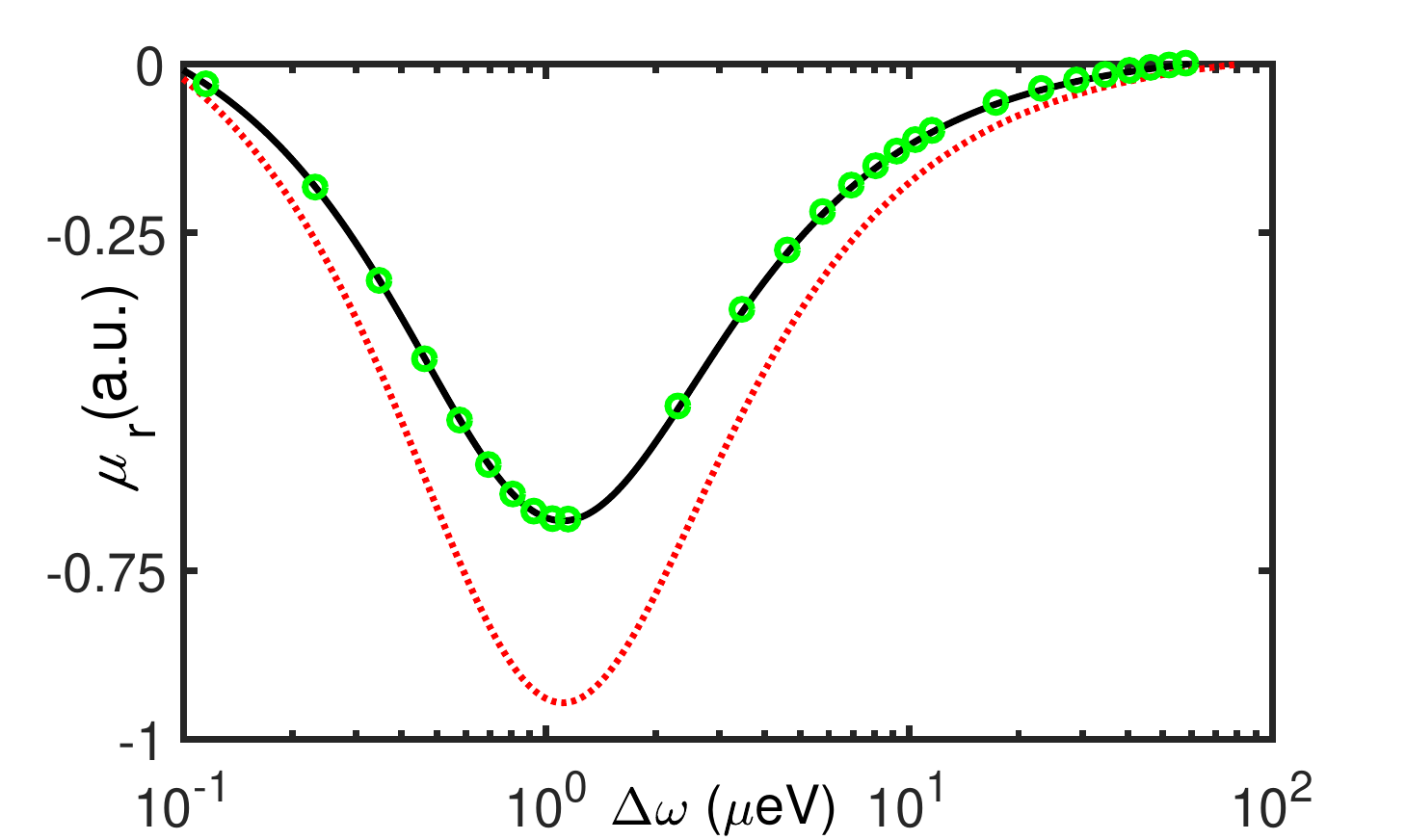}
\caption{ Effects of simultaneous multi-electron transitions and enhanced transition dipole on the relative permeability of a M\"{o}bius molecule.
The green circles is for
$\left\vert-\frac{\pi}{2},\downarrow\right\rangle\rightleftarrows\left\vert-\frac{\pi}{3},\uparrow\right\rangle$, and the black curve are for
$\left\vert\frac{\pi}{2},\downarrow\right\rangle\rightleftarrows\left\vert\frac{\pi}{2},\uparrow\right\rangle$,
and the red dotted curve sums up the two resonant transitions. }
\label{fig:Contribution}
\end{figure}

\textit{Conclusions}.\textendash{}\textendash{}We explore the possibility of broadening negative refraction by simultaneous multi-electron transitions. As a specific example, we calculate the optical properties of M\"{o}bius molecules based on the H\"{u}ckel molecular orbital method and linear response theory. The simultaneous multi-electron transitions provide parallel contributions to the overlapping negative refraction.
Furthermore, in each transition, the negative refraction is significantly broadened by the enhanced transition dipole. As a result of these two broadening effects,
the bandwidth for the negative refraction is larger than
the previous discoveries in Refs.~\cite{Kastel2007-1,Thommen2006}
by at least \textit{two} orders of magnitude.
Therefore, we clearly show that the contributions of multi-electron transitions
and enhanced transition dipoles can lead to a large broadening of the negative-index interval. Moreover, as metamaterials consisting of M\"{o}bius molecules are based on quantum transitions, their size is smaller by 2--3 orders of magnitude than in classical metamaterials.
Instead of producing a huge number of split-ring resonators,
metamaterials of M\"{o}bius molecules are chemically synthesized
and self-assembled by crystallization.
In conclusion, this work provides a possible route for broad-band negative
refractive-index of metamaterials.

\textit{Acknowledgments}.\textendash{}\textendash{}We thank
stimulating discussions with Y. N. Fang. This work was supported
by the National Natural Science Foundation of China
under Grant Nos.~11505007 and~11474026, the Fundamental Research Funds for the
Central Universities under Grant No.~2015KJJCA01,
MURI Center for Dynamic Magneto-Optics via the
Air Force Office of Scientific Research (AFOSR) (FA9550-14-1-0040),
Army Research Office (ARO) (Grant No.~73315PH),
Asian Office of Aerospace Research and Development (AOARD) (Grant No.~FA2386-18-1-4045),
Japan Science and Technology Agency (JST)
(the ImPACT program and CREST Grant No.~JPMJCR1676),
Japan Society for the Promotion of Science (JSPS) (JSPS-RFBR Grant No.~17-52-50023),
RIKEN-AIST Challenge Research Fund, and the
John Templeton Foundation.

J.-J.C. and Y.-Q.C. contributed equally to this work.

\title{Supplemental Material for \\``Broad-Band Negative Refraction via Simultaneous Multi-Electron Transitions''}

\author{Jing-Jing Cheng}

\address{Department of Physics, Applied Optics Beijing Area Major Laboratory,
Beijing Normal University, Beijing 100875, China}

\author{Ying-Qi Chu}

\address{Department of Physics, Applied Optics Beijing Area Major Laboratory,
Beijing Normal University, Beijing 100875, China}

\author{Tao Liu}

\address{Theoretical Quantum Physics Laboratory, RIKEN Cluster for Pioneering
Research, Wako-shi, Saitama 351-0198, Japan}

\author{Jie-Xing Zhao}

\address{Department of Physics, Applied Optics Beijing Area Major Laboratory,
Beijing Normal University, Beijing 100875, China}

\author{Fu-Guo Deng}

\address{Department of Physics, Applied Optics Beijing Area Major Laboratory,
Beijing Normal University, Beijing 100875, China}
\address{NAAM-Research Group, Department of Mathematics, Faculty of Science,
King Abdulaziz University, Jeddah 21589, Saudi Arabia}

\author{Qing Ai}
\email{aiqing@bnu.edu.cn}

\address{Theoretical Quantum Physics Laboratory, RIKEN Cluster for Pioneering
Research, Wako-shi, Saitama 351-0198, Japan}

\address{Department of Physics, Applied Optics Beijing Area Major Laboratory,
Beijing Normal University, Beijing 100875, China}

\author{Franco Nori}
\email{fnori@riken.jp}

\address{Theoretical Quantum Physics Laboratory, RIKEN Cluster for Pioneering
Research, Wako-shi, Saitama 351-0198, Japan}

\address{Department of Physics, The University of Michigan, Ann Arbor, Michigan
48109-1040, USA}

\date{\today}

\maketitle

\tableofcontents{}

\section{Multi-electron Transitions in M\"{o}bius Ring}

\label{sec:Model}

As shown in the main text, the single-electron Hamiltonian of a M\"{o}bius molecule
\begin{eqnarray}
H & = & \sum_{j=0}^{N-1}\left[A_{j}^{\dagger}MA_{j}-\xi\left(A_{j}^{\dagger}A_{j+1}+\mathrm{h.c.}\right)\right]
\end{eqnarray}
can be diagonalized as
\begin{equation}
H=C_{k}^{\dagger}E_{k}C_{k},
\end{equation}
where the eigenstates are
\begin{align}
\left|k,\uparrow\right\rangle   =\frac{1}{\sqrt{2N}}\sum_{j=0}^{N-1}\exp\left[-i\left(k-\frac{\delta}{2}\right)j\right]\left(a_{j}^{\dagger}-b_{j}^{\dagger}\right)\left|v\right\rangle ,\;
\left|k,\downarrow\right\rangle   =\frac{1}{\sqrt{2N}}\sum_{j=0}^{N-1}\exp(-ikj)\left(a_{j}^{\dagger}+b_{j}^{\dagger}\right)\left|v\right\rangle ,
\end{align}
with $\left|v\right\rangle $ the vacuum state. The corresponding eigenenergies are, respectively,
$E_{k\uparrow}=V-2\xi\cos\left(k-\delta/2\right)$
and $E_{k\downarrow}=-V-2\xi\cos k$.

In order to simulate the electromagnetic response of the M\"{o}bius
medium in the presence of applied fields, we employ linear-response
theory \cite{Kubo1985} to calculate the electric permittivity and magnetic permeability.
When there is an electric field applied on the M\"{o}bius molecule,
the molecule is polarized as
\begin{equation}
\langle\vec{d}\,\rangle=\int\frac{d\omega}{2\pi}S(\omega)\vec{E}(\omega)e^{-i\omega t},
\end{equation}
where
\begin{align}
\vec{E}(\omega) & =\int_{-\infty}^{\infty}dt\;\vec{E}(t)e^{i\omega t}
\end{align}
is the Fourier transform of the time-dependent electric field with
amplitude $\vec{E}{}_{0}$ and frequency $\omega$
\begin{align}
\vec{E}(t) &= \vec{E}{}_{0}\cos\omega t,\\
S(\omega) &= -J(\omega)-J^{\ast}(-\omega).
\end{align}
Here, the dipole-dipole correlation function reads
\begin{equation}
J(\omega)=-i\int_{0}^{\infty}dt\;\mathrm{\textrm{Tr}}\left[\vec{d}(t)\vec{d}\rho(0)\right]e^{i\omega t},
\end{equation}
where the initial state of the molecule is
$\rho(0)=\sum_in_{i}^{-1}\left\vert i\right\rangle \left\langle i\right\vert$ 
with $\sum_i n_{i}^{-1}=2N$ the total number of $\pi$-electrons. Particularly,
we obtain that $n_{i}=1$ when $\vert i\rangle=\vert k,\downarrow\rangle$ for ($-\frac{\pi}{2}\leq k\leq\frac{\pi}{2}$), denoted by the blue solid squares in Fig.~2 of the main text, or $\vert i\rangle=\vert k,\uparrow\rangle$ for ($-\frac{\pi}{6}\leq k\leq\frac{\pi}{3}$), denoted by the red solid circles in Fig.~2 of the main text; $n_{i}=2$ when $\vert i\rangle=\vert\pm\frac{2\pi}{3},\downarrow\rangle$, denoted by the green squares in Fig.~2 of the main text.

The electric dipole in the Heisenberg picture is
\begin{equation}
\vec{d}(t)=\exp(iH^{\dagger}t/\hbar)\;\vec{d}\;\exp(-iHt/\hbar).
\end{equation}
Because the Fourier transform of the electric field is
\begin{align}
\vec{E}(\omega_{1}) & =\pi\vec{E}_{0}[\delta(\omega_{1}+\omega)+\delta(\omega_{1}-\omega)],
\end{align}
the molecular electric dipole in the presence of an applied electric field is
\begin{eqnarray}
\langle\vec{d}\,\rangle = \int\frac{d\omega_{1}}{2\pi}S(\omega_{1})e^{-i\omega_{1}t}\pi\vec{E}_{0}[\delta(\omega_{1}+\omega)+\delta(\omega_{1}-\omega)]= -\vec{E}_{0}\mathrm{\textrm{Re}}\left\{\left[J(\omega)+J^{\ast}(-\omega)\right]e^{-i\omega t}\right\},
\end{eqnarray}
where
\begin{eqnarray}
J(\omega)\!\! = \!\!-i\int_{0}^{\infty}dt\;e^{i\omega t}\sum_{i}\frac{1}{n_{i}}\left\langle i\right|\vec{d}(t)\vec{d}\left|i\right\rangle \!\!= \!\! -i\int_{0}^{\infty}dt\;e^{i\omega t}\sum_{i}\frac{1}{n_{i}}\left\langle i\right|e^{iH^{\dagger}t/\hbar}\vec{d}e^{-iHt/\hbar}\vec{d}\left|i\right\rangle \!\!=\!\! \sum_{i,f}\frac{\vec{d}_{if}\vec{d}_{fi}}{n_{i}(\omega-\Delta_{fi}+i\gamma)}.\label{eq:J}
\end{eqnarray}
Here, in the sum, the final state should be different from the initial state, and
\begin{eqnarray}
\langle i\vert H\vert i\rangle & = & E_{i}-\frac{i}{2}\gamma
\end{eqnarray}
with $-i\gamma/2$ being phenomenologically introduced for the decay
of the excited state. In Eq.~(\ref{eq:J}), $\Delta_{fi}=(E_{f}-E_{i})/\hbar$
is the transition frequency between the initial state $\left\vert i\right\rangle$ and the final state $\left\vert f\right\rangle $. Therefore, the molecular electric
dipole can be rewritten as
\begin{eqnarray}
\langle\vec{d}\,\rangle\!\! & \!=\! & \!\!-\sum_{i,f}\frac{\vec{E}_{0}}{2n_i}\left\{ \left[\left(\frac{\vec{d}_{if}\vec{d}_{fi}}{\omega-\Delta_{fi}+i\gamma}-\frac{\vec{d}_{if}\vec{d}_{fi}}{\omega+\Delta_{fi}+i\gamma}\right)e^{-i\omega t}\right.\right.+\left.\left.\left(\frac{\vec{d}_{if}\vec{d}_{fi}}{\omega-\Delta_{fi}-i\gamma}-\frac{\vec{d}_{if}\vec{d}_{fi}}{\omega+\Delta_{fi}-i\gamma}\right)e^{i\omega t}\right]\right\} .
\end{eqnarray}
Because of the rotating-wave approximation \cite{Ai2010}, the second
and third terms of the above equation can be ignored as
\begin{eqnarray}
\langle\vec{d}\,\rangle \approx  -\sum_{i,f}\frac{\vec{E}_{0}}{2n_{i}}\left(\frac{\vec{d}_{if}\vec{d}_{fi}e^{-i\omega t}}{\omega-\Delta_{fi}+i\gamma}+\frac{\vec{d}_{if}\vec{d}_{fi}e^{i\omega t}}{\omega-\Delta_{fi}-i\gamma}\right)  \approx -\sum_{i,f}\textrm{Re}\left[\frac{\vec{d}_{if}\vec{d}_{fi}\cdot\vec{E}(t)}{ n_{i}\hbar\left(\omega-\Delta_{fi}+i\gamma\right) }\right].
\end{eqnarray}

Assuming that all molecules are identical, the polarization density
reads
\begin{equation}
\vec{P}=\frac{\langle\vec{d}\,\rangle}{\hbar v_{0}},
\end{equation}
where $v_{0}\simeq2\pi(R+W)^{2}W$ is the volume of the molecule.

Because the electric displacement field is \cite{Landau1995,Jackson1999}
\begin{eqnarray}
\vec{D} & = & \varepsilon_{0}\vec{E}+\vec{P}=\varepsilon_{0}\overleftrightarrow{\varepsilon_{r}}\vec{E},\label{eq:D}
\end{eqnarray}
the relative permittivity is explicitly given as
\begin{equation}
\overleftrightarrow{\varepsilon_{r}}=1-\sum_{i,f}\frac{\vec{d}_{if}\vec{d}_{fi}}{n_{i}\hbar\varepsilon_{0}v_{0}}\textrm{Re}\left(\frac{1}{\omega-\Delta_{fi}+i\gamma}\right).\label{eq:Permit}
\end{equation}
In a similar way, the relative permeability is calculated as
\begin{equation}
\overleftrightarrow{\mu_{r}}=1-\sum_{i,f}\frac{\mu_{0}\vec{m}{}_{if}\vec{m}_{fi}}{n_{i}\hbar v_{0}}\textrm{Re}\left(\frac{1}{\omega-\Delta_{fi}+i\gamma}\right).\label{eq:Permea}
\end{equation}

\section{Selection Rules for Transitions }

In the previous section, we apply linear response theory to rewrite
the relative permittivity and permeability in terms of the transition
matrix elements of the electric and magnetic dipole, i.e.,
$\vec{d}{}_{if}$ and $\vec{m}{}_{if}$. According to
Eqs.~(\ref{eq:Permit},\ref{eq:Permea}), we would expect the negative
permittivity and permeability to be close to the transition frequency $\Delta_{fi}$.
Hereafter, the selection rules of transitions are explicitly provided
by calculating the matrix elements of the electric and magnetic dipole.

Under the dipole approximation, the interaction Hamiltonian with the electric
field is written as \cite{Landau1995,Jackson1999}
\begin{equation}
H_{E}^{'}=-\vec{d}\cdot\vec{E}_{0}\textrm{cos}\omega t.
\end{equation}
The electric-dipole-induced transition is allowed if the corresponding
matrix element is nonzero. By using the relation $\left\langle \phi_{js}\right|\vec{r}\left|\phi_{j's'}\right\rangle =\vec{R}_{js}\delta_{jj^{\prime}}\delta_{ss^{\prime}}$ \cite{Salem1972},
the non-vanishing matrix elements are explicitly listed as
\begin{subequations}\label{eq:HE}
\begin{eqnarray}
\left\langle 0\right|C_{k\uparrow}^{\dagger}H_{E}^{'}C_{k\downarrow}\left|0\right\rangle \!\! & \!=\! & \!\!\left(i\frac{E_{0}^{(x)}}{4}+\frac{E_{0}^{(y)}}{4}+\frac{E_{0}^{(z)}}{2}\right)eW\textrm{cos}\omega t, \\
\left\langle 0\right|C_{k\downarrow}^{\dagger}H_{E}^{'}C_{k\uparrow}\left|0\right\rangle \!\! & = & \!\!\left(-i\frac{E_{0}^{(x)}}{4}+\frac{E_{0}^{(y)}}{4}+\frac{E_{0}^{(z)}}{2}\right)eW\textrm{cos}\omega t, \\
\left\langle 0\right|C_{k\uparrow}^{\dagger}H_{E}^{'}C_{k+\delta,\uparrow}\left|0\right\rangle \!\! & = & \!\!\left(\frac{E_{0}^{(x)}}{2}-i\frac{E_{0}^{(y)}}{2}\right)eR\textrm{cos}\omega t, \\
\left\langle 0\right|C_{k\uparrow}^{\dagger}H_{E}^{'}C_{k+\delta,\downarrow}\left|0\right\rangle \!\! & = & \!\!\left(-i\frac{E_{0}^{(x)}}{4}-\frac{E_{0}^{(y)}}{4}\right)eW\textrm{cos}\omega t, \\
\left\langle 0\right|C_{k\downarrow}^{\dagger}H_{E}^{'}C_{k+\delta,\uparrow}\left|0\right\rangle \!\! & = & \!\!\left(i\frac{E_{0}^{(x)}}{4}+\frac{E_{0}^{(y)}}{4}+\frac{E_{0}^{(z)}}{2}\right)eW\textrm{cos}\omega t, \\
\left\langle 0\right|C_{k\downarrow}^{\dagger}H_{E}^{'}C_{k+\delta,\downarrow}\left|0\right\rangle \!\! & = & \!\!\left(\frac{E_{0}^{(x)}}{2}-i\frac{E_{0}^{(y)}}{2}\right)eR\textrm{cos}\omega t,\\
\left\langle 0\right|C_{k\uparrow}^{\dagger}H_{E}^{'}C_{k-\delta,\uparrow}\left|0\right\rangle \!\! & = & \!\!\left(\frac{E_{0}^{(x)}}{2}+i\frac{E_{0}^{(y)}}{2}\right)eR\textrm{cos}\omega t, \\
\left\langle 0\right|C_{k\uparrow}^{\dagger}H_{E}^{'}C_{k-\delta,\downarrow}\left|0\right\rangle \!\! & = & \!\!\left(-i\frac{E_{0}^{(x)}}{4}+\frac{E_{0}^{(y)}}{4}+\frac{E_{0}^{(z)}}{2}\right)eW\textrm{cos}\omega t, \\
\left\langle 0\right|C_{k\downarrow}^{\dagger}H_{E}^{'}C_{k-\delta,\uparrow}\left|0\right\rangle \!\! & = & \!\!\left(i\frac{E_{0}^{(x)}}{4}-\frac{E_{0}^{(y)}}{4}\right)eW\textrm{cos}\omega t, \\
\left\langle 0\right|C_{k\downarrow}^{\dagger}H_{E}^{'}C_{k-\delta,\downarrow}\left|0\right\rangle \!\! & = & \!\!\left(\frac{E_{0}^{(x)}}{2}+i\frac{E_{0}^{(y)}}{2}\right)eR\textrm{cos}\omega t, \\
\left\langle 0\right|C_{k\downarrow}^{\dagger}H_{E}^{'}C_{k+2\delta,\uparrow}\left|0\right\rangle \!\! & = & \!\!\left(-i\frac{E_{0}^{(x)}}{4}-\frac{E_{0}^{(y)}}{4}\right)eW\textrm{cos}\omega t.
\end{eqnarray}
\end{subequations}

For example, because in Eq.~(\ref{eq:HE}), the matrix element is nonzero between
the states $\left|k,\uparrow\right\rangle$ and $\left|k,\downarrow\right\rangle$,
and there are contributions from all three components of the electric field, the transition $\left|k,\uparrow\right\rangle \rightleftharpoons\left|k,\downarrow\right\rangle$
can be electric-dipole induced for all three polarizations of the
electric field. Based on Eq.~(\ref{eq:HE}), the selection rules for the
electric-dipole induced transitions are summarized as
\begin{eqnarray}\label{eq:ElecSelRule}
\left|k,\downarrow\right\rangle   \overset{x,y,z}{\rightleftharpoons}  \left|k,\uparrow\right\rangle ,\;
\left|k,\downarrow\right\rangle   \overset{x,y}{\rightleftharpoons}  \left|k+2\delta,\uparrow\right\rangle ,\;
\left|k,\downarrow\right\rangle   \overset{x,y,z}{\rightleftharpoons}  \left|k+\delta,\uparrow\right\rangle ,\;
\left|k,\downarrow\right\rangle   \overset{x,y}{\rightleftharpoons}  \left|k-\delta,\uparrow\right\rangle ,
\end{eqnarray}
where the superscripts over the arrows indicate
the polarization of the electric field.

Under the dipole approximation, the interaction Hamiltonian with the magnetic field
is written as \cite{Landau1995,Jackson1999}
\begin{equation}
H_{B}^{'}=-\vec{m}\cdot\vec{B}_{0}\textrm{cos}\omega t.
\end{equation}
Straightforward calculations show the following
non-vanishing matrix elements,
\begin{subequations}\label{eq:HB1}
\begin{eqnarray}
\left\langle 0\right|C_{k\uparrow}^{\dagger}H_{B}^{'}C_{k\uparrow}\left|0\right\rangle  & = & -\left\{ 2W^{2}\left(\textrm{cos}(k-\delta)-\textrm{cos}k\right)B_{0}^{(y)}+\left[W^{2}\left(\textrm{cos}k-\textrm{cos}(k-2\delta)-\textrm{cos}(k-\delta)+\textrm{cos}(k+\delta)\right)\right.\right. \nonumber\\
 &  & \left.\left.+4R^{2}\left(\textrm{cos}(k+\frac{\delta}{2})-\textrm{cos}(k-\frac{3}{2}\delta)\right)\right]B_{0}^{(z)}\right\} \frac{e\xi}{8\hbar}\textrm{cos}\omega t,
 \\
\left\langle 0\right|C_{k\uparrow}^{\dagger}H_{B}^{'}C_{k\downarrow}\left|0\right\rangle  & = & -\left\{ \left[V+\xi\left(\textrm{cos}(k-\delta)-\textrm{cos}(k+\frac{\delta}{2})\right)\right]\left(B_{0}^{(x)}-iB_{0}^{(y)}\right)\right. \nonumber\\
 &  & \left.+2i\xi\textrm{cos}\frac{\delta}{4}\left(\textrm{cos}(k-\frac{5\delta}{4})-\textrm{cos}(k+\frac{3}{4}\delta)\right)B_{0}^{(z)}\right\} \frac{eRW}{4\hbar}\textrm{cos}\omega t,
 \\
\left\langle 0\right|C_{k\downarrow}^{\dagger}H_{B}^{'}C_{k\uparrow}\left|0\right\rangle  & = & -\left\{ \left[V+\xi\left(\textrm{cos}(k-\delta)-\textrm{cos}(k+\frac{\delta}{2})\right)\right]\left(B_{0}^{(x)}+iB_{0}^{(y)}\right)\right. \nonumber\\
 &  & \left.+2i\xi\textrm{cos}\frac{\delta}{4}\left(\textrm{cos}(k+\frac{3\delta}{4})-\textrm{cos}(k-\frac{5}{4}\delta)\right)B_{0}^{(z)}\right\} \frac{eRW}{4\hbar}\textrm{cos}\omega t,
 \\
\left\langle 0\right|C_{k\downarrow}^{\dagger}H_{B}^{'}C_{k\downarrow}\left|0\right\rangle  & = & -\left[W^{2}B_{0}^{(y)}\textrm{sin}\frac{\delta}{2}-\left(2R^{2}+W^{2}\textrm{cos}\frac{\delta}{2}\right)B_{0}^{(z)}\textrm{sin}\delta\right]\frac{e\xi}{2\hbar}\textrm{sin}k\textrm{cos}\omega t, \\
\left\langle 0\right|C_{k\uparrow}^{\dagger}H_{B}^{'}C_{k+\delta,\uparrow}\left|0\right\rangle  & = & \left(\textrm{cos}(k-\delta)-\textrm{cos}(k+\delta)\right)\left(iB_{0}^{(x)}+B_{0}^{(y)}-B_{0}^{(z)}\right)\frac{eW^{2}\xi}{8\hbar}\textrm{cos}\omega t, \\
\left\langle 0\right|C_{k\uparrow}^{\dagger}H_{B}^{'}C_{k+\delta,\downarrow}\left|0\right\rangle  & = & -\left\{ \left[V+\xi\left(\textrm{cos}k-\textrm{cos}(k+\frac{\delta}{2})\right)\right]\left(B_{0}^{(x)}-iB_{0}^{(y)}\right)\right. \nonumber\\
 &  & \left.-2i\xi\textrm{cos}\frac{\delta}{4}\left(\textrm{cos}(k-\frac{5\delta}{4})-\textrm{cos}(k+\frac{3}{4}\delta)\right)B_{0}^{(z)}\right\} \frac{eRW}{4\hbar}\textrm{cos}\omega t,
\\
\left\langle 0\right|C_{k\downarrow}^{\dagger}H_{B}^{'}C_{k+\delta,\uparrow}\left|0\right\rangle  & = & \left\{ \left[V+\xi\left(\textrm{cos}(k+\delta)-\textrm{cos}(k-\frac{\delta}{2})\right)\right]\left(B_{0}^{(x)}-iB_{0}^{(y)}\right)\right. \nonumber\\
 &  & \left.-i\xi\left(\textrm{cos}(k-\delta)-\textrm{cos}(k+\delta)-\textrm{cos}(k+\frac{3\delta}{2})+\textrm{cos}(k-\frac{\delta}{2})\right)B_{0}^{(z)}\right\} \frac{eRW}{4\hbar}\textrm{cos}\omega t,
 \\
\left\langle 0\right|C_{k\downarrow}^{\dagger}H_{B}^{'}C_{k+\delta,\downarrow}\left|0\right\rangle  & = & \left(\textrm{cos}(k-\frac{\delta}{2})-\textrm{cos}(k+\frac{3\delta}{2})\right)\left(iB_{0}^{(x)}+B_{0}^{(y)}-B_{0}^{(z)}\right)\frac{eW^{2}\xi}{8\hbar}\textrm{cos}\omega t,\\
\left\langle 0\right|C_{k\uparrow}^{\dagger}H_{B}^{'}C_{k-\delta,\uparrow}\left|0\right\rangle  & = & -\left(\textrm{cos}(k-2\delta)-\textrm{cos}k\right)\left(iB_{0}^{(x)}-B_{0}^{(y)}+B_{0}^{(z)}\right)\frac{eW^{2}\xi}{8\hbar}\textrm{cos}\omega t,
\\
\left\langle 0\right|C_{k\uparrow}^{\dagger}H_{B}^{'}C_{k-\delta,\downarrow}\left|0\right\rangle  & = & \left\{ \left[V+\xi\left(\textrm{cos}k-\textrm{cos}(k-\frac{3\delta}{2})\right)\right]\left(B_{0}^{(x)}+iB_{0}^{(y)}\right)\right. \nonumber\\
 &  & \left.+i\xi\left(\textrm{cos}(k-\frac{3\delta}{2})+\textrm{cos}(k-2\delta)-\textrm{cos}k-\textrm{cos}(k+\frac{\delta}{2})\right)B_{0}^{(z)}\right\} \frac{eRW}{4\hbar}\textrm{cos}\omega t,
 \\
\left\langle 0\right|C_{k\downarrow}^{\dagger}H_{B}^{'}C_{k-\delta,\uparrow}\left|0\right\rangle  & = & -\left[V+\xi\left(\textrm{cos}(k-\delta)-\textrm{cos}(k-\frac{\delta}{2})\right)\right]\left(B_{0}^{(x)}+iB_{0}^{(y)}\right)\frac{eRW}{4\hbar}\textrm{cos}\omega t,\\
\left\langle 0\right|C_{k\downarrow}^{\dagger}H_{B}^{'}C_{k-\delta,\downarrow}\left|0\right\rangle  & = & \left(\textrm{cos}(k-\frac{3\delta}{2})-\textrm{cos}(k+\frac{\delta}{2})\right)\left(-iB_{0}^{(x)}+B_{0}^{(y)}-B_{0}^{(z)}\right)\frac{eW^{2}\xi}{8\hbar}\textrm{cos}\omega t, \\
\left\langle 0\right|C_{k\uparrow}^{\dagger}H_{B}^{'}C_{k+2\delta,\uparrow}\left|0\right\rangle  & = & i\left(\textrm{cos}k-\textrm{cos}(k+\delta)\right)\left(B_{0}^{(x)}-iB_{0}^{(y)}\right)\frac{eW^{2}\xi}{8\hbar}\textrm{cos}\omega t, \\
\left\langle 0\right|C_{k\downarrow}^{\dagger}H_{B}^{'}C_{k+2\delta,\uparrow}\left|0\right\rangle  & = & \left[V+\xi\left(\textrm{cos}(k+\delta)-\textrm{cos}(k+\frac{\delta}{2})\right)\right]\left(B_{0}^{(x)}-iB_{0}^{(y)}\right)\frac{eRW}{4\hbar}\textrm{cos}\omega t, \\
\left\langle 0\right|C_{k\downarrow}^{\dagger}H_{B}^{'}C_{k+2\delta,\downarrow}\left|0\right\rangle  & = & i\left(\textrm{cos}(k+\frac{\delta}{2})-\textrm{cos}(k+\frac{3\delta}{2})\right)\left(B_{0}^{(x)}-iB_{0}^{(y)}\right)\frac{eW^{2}\xi}{8\hbar}\textrm{cos}\omega t.
\end{eqnarray}
\end{subequations}
Therefore, we have the following selection rules for the magnetic-dipole-induced transitions, i.e.,
\begin{eqnarray}\label{eq:MagSelRule}
\left|k,\downarrow\right\rangle   \overset{x,y,z}{\rightleftharpoons}  \left|k,\uparrow\right\rangle ,\;
\left|k,\downarrow\right\rangle   \overset{x,y}{\rightleftharpoons}  \left|k+2\delta,\uparrow\right\rangle ,\;
\left|k,\downarrow\right\rangle   \overset{x,y,z}{\rightleftharpoons}  \left|k+\delta,\uparrow\right\rangle ,\;
\left|k,\downarrow\right\rangle   \overset{x,y}{\rightleftharpoons}  \left|k-\delta,\uparrow\right\rangle ,
\end{eqnarray}
which is the same as the selection rules (\ref{eq:ElecSelRule}) for the electric-dipole
induced transitions.

Because a transition can take place when there is an electron in the highest occupied molecular
orbital (HOMO), and the lowest unoccupied molecular orbital (LUMO)
is unoccupied. The transitions allowed by both electric and magnetic dipole couplings
are schematically shown in Fig.~2 of the main text.

\section{Permittivity}

\label{app:Permittivity}

According to the selection rules of optical transitions in M\"{o}bius
molecules, there are four possible transitions from a given initial state.
Generally speaking, the permittivity is a second-rank tensor with
nine matrix elements, i.e.,
\begin{equation}
\overleftrightarrow{\varepsilon_{r}}=\sum_{i,j=x,y,z}\varepsilon_{r}^{i,j}\hat{e}_{i}\hat{e}_{j}.\label{eq:Er}
\end{equation}
According to Eqs.~(\ref{eq:Permit},\ref{eq:HE}), the three diagonal
components can be calculated as
\begin{subequations}
\begin{eqnarray}
\varepsilon_{r}^{xx}
 & = & 1-\sum_{k}\left(\eta_{k,k}^{\prime}+\eta_{k,k+\delta}^{\prime}+\eta_{k,k+2\delta}^{\prime}+\eta_{k,k-\delta}^{\prime}\right),\label{eq:exx}\\
\varepsilon_{r}^{yy}
 & = & 1-\sum_{k}\left(\eta_{k,k}^{\prime}+\eta_{k,k+\delta}^{\prime}+\eta_{k,k+2\delta}^{\prime}+\eta_{k,k-\delta}^{\prime}\right),\label{eq:eyy}\\
\varepsilon_{r}^{zz}
 & = & 1-\sum_{k}\left(4\eta_{k,k}^{\prime}+4\eta_{k,k+\delta}^{\prime}\right),\label{eq:ezz}
\end{eqnarray}
\end{subequations}
where $\eta^{\prime}_{k,k'}(\omega)$ is the real part of $\eta_{k,k'}(\omega)  =\eta^{\prime}_{k,k'}(\omega)+i\eta^{\prime\prime}_{k,k'}(\omega)$,
\begin{subequations}
\begin{align}
\eta_{k,k} & =\frac{1}{16\hbar\varepsilon_{0}v_{0}}\frac{e^{2}W^{2}}{\omega-2V-2\xi\left[\cos k-\cos\left(k-\frac{\delta}{2}\right)\right]+i\gamma},\\
\eta_{k,k+\delta} & =\frac{1}{16\hbar\varepsilon_{0}v_{0}}\frac{e^{2}W^{2}}{\omega-2V-2\xi\left[\cos k-\cos\left(k+\frac{\delta}{2}\right)\right]+i\gamma},\\
\eta_{k,k+2\delta} & =\frac{1}{16\hbar\varepsilon_{0}v_{0}}\frac{e^{2}W^{2}}{\omega-2V-2\xi\left[\cos k-\cos\left(k+\frac{3\delta}{2}\right)\right]+i\gamma},\\
\eta_{k,k-\delta} & =\frac{1}{16\hbar\varepsilon_{0}v_{0}}\frac{e^{2}W^{2}}{\omega-2V-2\xi\left[\cos k-\cos\left(k-\frac{3\delta}{2}\right)\right]+i\gamma}.
\end{align}
\end{subequations}
Notice that $\eta^{\prime}_{k,k'}$ are of the same form but with a different central frequency.
The six off-diagonal matrix elements of the relative permittivity
read\begin{subequations}
\begin{align}
\varepsilon_{r}^{xy} & =-\varepsilon_{r}^{yx} =\sum_{k}i\left(\eta_{k,k}^{\prime}-\eta_{k,k+\delta}^{\prime}-\eta_{k,k+2\delta}^{\prime}+\eta_{k,k-\delta}^{\prime}\right),\\
\varepsilon_{r}^{xz} & =-\varepsilon_{r}^{zx} =\sum_{k}2i\left(\eta_{k,k}^{\prime}-\eta_{k,k+\delta}^{\prime}\right),\\
\varepsilon_{r}^{yz} & =\varepsilon_{r}^{zy} =-\sum_{k}2\left(\eta_{k,k}^{\prime}+\eta_{k,k+\delta}^{\prime}\right).
\end{align}
\end{subequations}

In addition to the selection rules of optical transitions, the electric
response can only occur at those transitions from the HOMOs to the LUMOs.
The relative permittivity $\overleftrightarrow{\varepsilon_{r}}$
can be further simplified as
\begin{equation}
\overleftrightarrow{\varepsilon_{r}}=1+\sum_{k\in\mathrm{HOMO}}\;\sum_{k'\in\mathrm{LUMO}}\overleftrightarrow{\chi}_{k,k'}^{\textrm{E}},\label{eq:final:e}
\end{equation}
where the second term can be explicitly given as
\begin{align}
\sum_{k\in\mathrm{HOMO}}\;\sum_{k'\in\mathrm{LUMO}}\overleftrightarrow{\chi}_{k,k'}^{\textrm{E}}
=&\frac{1}{2}  \left(\overleftrightarrow{\chi}_{-\frac{2\pi}{3},-\frac{5\pi}{6}}^{\textrm{E}}+\overleftrightarrow{\chi}_{-\frac{2\pi}{3},-\frac{2\pi}{3}}^{\textrm{E}}+\overleftrightarrow{\chi}_{-\frac{2\pi}{3},-\frac{\pi}{2}}^{\textrm{E}}+\overleftrightarrow{\chi}_{-\frac{2\pi}{3},-\frac{\pi}{3}}^{\textrm{E}}\right) +\overleftrightarrow{\chi}_{-\frac{\pi}{2},-\frac{2\pi}{3}}^{\textrm{E}}+\overleftrightarrow{\chi}_{-\frac{\pi}{2},-\frac{\pi}{2}}^{\textrm{E}}\nonumber\\
&+\overleftrightarrow{\chi}_{-\frac{\pi}{2},-\frac{\pi}{3}}^{\textrm{E}}+\overleftrightarrow{\chi}_{-\frac{\pi}{3},-\frac{2\pi}{3}}^{\textrm{E}} +\overleftrightarrow{\chi}_{-\frac{\pi}{3},-\frac{\pi}{3}}^{\textrm{E}}+\overleftrightarrow{\chi}_{-\frac{\pi}{6},-\frac{\pi}{3}}^{\textrm{E}}+\overleftrightarrow{\chi}_{\frac{\pi}{6},\frac{\pi}{2}}^{\textrm{E}}+\overleftrightarrow{\chi}_{\frac{\pi}{3},\frac{\pi}{2}}^{\textrm{E}} +\overleftrightarrow{\chi}_{\frac{\pi}{3},\frac{2\pi}{3}}^{\textrm{E}}\nonumber \\
 &+\overleftrightarrow{\chi}_{\frac{\pi}{2},\frac{\pi}{2}}^{\textrm{E}}+\overleftrightarrow{\chi}_{\frac{\pi}{2},\frac{2\pi}{3}}^{\textrm{E}}+\overleftrightarrow{\chi}_{\frac{\pi}{2},\frac{\text{5\ensuremath{\pi}}}{6}}^{\textrm{E}} +\frac{1}{2}\left(\overleftrightarrow{\chi}_{\frac{2\pi}{3},\frac{\pi}{2}}^{\textrm{E}}+\overleftrightarrow{\chi}_{\frac{2\pi}{3},\frac{2\pi}{3}}^{\textrm{E}}+\overleftrightarrow{\chi}_{\frac{2\pi}{3},\frac{5\pi}{6}}^{\textrm{E}}+\overleftrightarrow{\chi}_{\frac{2\pi}{3},\pi}^{\textrm{E}}\right).
\end{align}
In the first and third lines, the factors $1/2$ are due to half occupation of the levels at $\vert \pm\frac{2\pi}{3},\uparrow\rangle$. The components of the tensor
\begin{equation}
\overleftrightarrow{\chi}_{k,k'}^{\textrm{E}}=\sum_{i,j=x,y,z}\chi_{k,k'}^{\mathrm{E},i,j}\hat{e}_{i}\hat{e}_{j}
\end{equation}
are respectively
\begin{subequations}
\begin{align}
\text{\ensuremath{\chi}}_{k,k'}^{\mathrm{E},xx} & =\text{\ensuremath{\chi}}_{k,k'}^{\mathrm{E},yy}=-\eta_{k,k'}^{\prime},\\
\text{\ensuremath{\chi}}_{k,k'}^{\mathrm{E},zz} & =-4\eta_{k,k'}^{\prime},\\
\text{\ensuremath{\chi}}_{k,k'}^{\mathrm{E},xy} & =-\text{\ensuremath{\chi}}_{k,k'}^{\mathrm{E},xy} =i\eta_{k,k'}^{\prime}(\delta{}_{k',k}-\delta{}_{k',k+\delta}-\delta_{k',k+2\delta}+\delta_{k',k-\delta}),\\
\text{\ensuremath{\chi}}_{k,k'}^{\mathrm{E},xz} & =-\text{\ensuremath{\chi}}_{k,k'}^{\mathrm{E},zx}=i2\eta_{k,k'}^{\prime}(\delta{}_{k',k}-\delta{}_{k',k+\delta}),\\
\text{\ensuremath{\chi}}_{k,k'}^{\mathrm{E},yz} & =\text{\ensuremath{\chi}}_{k,k'}^{\mathrm{E},zy}=-2\eta_{k,k'}^{\prime}(\delta_{k',k+\delta}+\delta_{k',k}).
\end{align}
\end{subequations}

\section{Permeability}

\label{app:Permeability}

Following the same procedure, we can calculate the nine components
of the relative permeability, i.e.,
\begin{equation}
\overleftrightarrow{\mu_{r}}=\sum_{i,j=x,y,z}\mu_{r}^{i,j}\hat{e}_{i}\hat{e}_{j}.\label{eq:Mu}
\end{equation}
According to Eqs.~(\ref{eq:Permea},\ref{eq:HB1}), the three diagonal
components of the relative permeability are respectively
\begin{subequations}
\begin{eqnarray}
\mu_{r}^{xx} & = & 1-\sum_{k}\left(\alpha_{k,k}^{2}\eta_{k,k}^{\prime}+\alpha_{k,k+\delta}^{2}\eta_{k,k+\delta}^{\prime} +\alpha_{k,k+2\delta}^{2}\eta_{k,k+2\delta}^{\prime}+\alpha_{k,k-\delta}^{2}\eta_{k,k-\delta}^{\prime}\right),\\
\mu_{r}^{yy} & = & 1-\sum_{k}\left(\alpha_{k,k}^{2}\eta_{k,k}^{\prime}+\alpha_{k,k+\delta}^{2}\eta_{k,k+\delta}^{\prime} +\alpha_{k,k+2\delta}^{2}\eta_{k,k+2\delta}^{\prime}+\alpha_{k,k-\delta}^{2}\eta_{k,k-\delta}^{\prime}\right),\\
\mu_{r}^{zz} & = & 1-\sum_{k}\left(4\beta_{k,k}^{2}\eta_{k,k}^{\prime}+4\beta_{k,k+\delta}^{2}\eta_{k,k+\delta}^{\prime}\right),
\end{eqnarray}
\end{subequations}
where
\begin{subequations}
\begin{eqnarray}
\alpha_{k,k}\!\! & = & \!\!\frac{R}{\hbar c}\left\{ V+\varepsilon\left[\textrm{cos}(k-\delta)-\textrm{cos}\left(k+\frac{\delta}{2}\right)\right]\right\} =\frac{R}{\hbar c}\left[V+2\varepsilon\sin\left(k-\frac{\delta}{2}\right)\sin\frac{3\delta}{4}\right],\\
\alpha_{k,k+\delta}\!\! & = & \!\!\frac{R}{\hbar c}\left\{ V+\varepsilon\left[\textrm{cos}(k+\delta)-\textrm{cos}\left(k-\frac{\delta}{2}\right)\right]\right\} =\frac{R}{\hbar c}\left[V-2\varepsilon\sin\left(k-\frac{\delta}{2}\right)\sin\frac{3\delta}{4}\right],\\
\alpha_{k,k+2\delta}\!\! & = & \!\!\frac{R}{\hbar c}\left\{ V+\varepsilon\left[\textrm{cos}(k+\delta)-\textrm{cos}\left(k+\frac{\delta}{2}\right)\right]\right\} =\frac{R}{\hbar c}\left[V-2\varepsilon\sin\left(k+\frac{3\delta}{4}\right)\sin\frac{\delta}{4}\right],\\
\alpha_{k,k-\delta}\!\! & = & \!\!\frac{R}{\hbar c}\left\{ V+\varepsilon\left[\textrm{cos}(k-\delta)-\textrm{cos}\left(k-\frac{\delta}{2}\right)\right]\right\} =\frac{R}{\hbar c}\left[V+2\varepsilon\sin\left(k-\frac{3\delta}{4}\right)\sin\frac{\delta}{4}\right],\\
\beta_{k,k}\!\! & = & \!\!\frac{R\varepsilon}{\hbar c}\textrm{cos}\frac{\delta}{4}\left[\textrm{cos}\left(k+\frac{3\delta}{4}\right)-\textrm{cos}\left(k-\frac{5\delta}{4}\right)\right]
=-\frac{2R\varepsilon}{\hbar c}\sin\left(k-\frac{\delta}{4}\right)\sin\delta\cos\frac{\delta}{4},\\
\beta_{k,k+\delta}\!\! & = & \!\!\frac{R\varepsilon}{2\hbar c}\left[\textrm{cos}(k-\delta)-\textrm{cos}(k+\delta)-\textrm{cos}\left(k+\frac{3\delta}{2}\right)\right.\left.+\textrm{cos}\left(k-\frac{\delta}{2}\right)\right]
=\frac{2R\varepsilon}{\hbar c}\sin\left(k+\frac{\delta}{4}\right)\sin\delta\cos\frac{\delta}{4}.
\end{eqnarray}
\end{subequations}
The six off-diagonal components are respectively
\begin{subequations}
\begin{eqnarray}
\mu_{r}^{xy}  &  =  &  -\mu_{r}^{yx}= \sum_{k}i\left(\alpha_{k,k}^{2}\eta_{k,k}^{\prime}-\alpha_{k,k+\delta}^{2}\eta_{k,k+\delta}^{\prime} -\alpha_{k,k+2\delta}^{2}\eta_{k,k+2\delta}^{\prime}+\alpha_{k,k-\delta}^{2}\eta_{k,k-\delta}^{\prime}\right),\\
\mu_{r}^{xz}  &  =  &  -\mu_{r}^{zx}= \sum_{k}2i\left(\alpha_{k,k}\beta_{k,k}\eta_{k,k}^{\prime}-\alpha_{k,k+\delta}\beta_{k,k+\delta}\eta_{k,k+\delta}^{\prime}\right),\\
\mu_{r}^{yz}  &  =  &  \mu_{r}^{zy}= -\sum_{k}2\left(\alpha_{k,k}\beta_{k,k}\eta_{k,k}^{\prime}+\alpha_{k,k+\delta}\beta_{k,k+\delta}\eta_{k,k+\delta}^{\prime}\right).
\end{eqnarray}
\end{subequations}

In addition to the selection rules of optical transitions, the magnetic
response can only occur at those transitions from the HOMOs to the LUMOs.
The relative permittivity $\overleftrightarrow{\mu_{r}}$
are related to $\overleftrightarrow{\chi}_{k,k'}^{\textrm{B}}$ by
\begin{equation}
\overleftrightarrow{\mu_{r}}=1+\sum_{k\in\mathrm{HOMO}}\;\sum_{k'\in\mathrm{LUMO}}\overleftrightarrow{\chi}_{k,k'}^{\textrm{B}},\label{eq:final:mu}
\end{equation}
where
\begin{align}
 \sum_{k\in\mathrm{HOMO}}\;\sum_{k'\in\mathrm{LUMO}}\overleftrightarrow{\chi}_{k,k'}^{\textrm{B}}
= & \frac{1}{2} \left(\overleftrightarrow{\chi}_{-\frac{2\pi}{3},-\frac{5\pi}{6}}^{\textrm{B}}+\overleftrightarrow{\chi}_{-\frac{2\pi}{3},-\frac{2\pi}{3}}^{\textrm{B}}+\overleftrightarrow{\chi}_{-\frac{2\pi}{3},-\frac{\pi}{2}}^{\textrm{B}}+\overleftrightarrow{\chi}_{-\frac{2\pi}{3},-\frac{\pi}{3}}^{\textrm{B}}\right) +\overleftrightarrow{\chi}_{-\frac{\pi}{2},-\frac{2\pi}{3}}^{\textrm{B}}+\overleftrightarrow{\chi}_{-\frac{\pi}{2},-\frac{\pi}{2}}^{\textrm{B}}\nonumber \\
 &+\overleftrightarrow{\chi}_{-\frac{\pi}{2},-\frac{\pi}{3}}^{\textrm{B}}+\overleftrightarrow{\chi}_{-\frac{\pi}{3},-\frac{2\pi}{3}}^{\textrm{B}} +\overleftrightarrow{\chi}_{-\frac{\pi}{3},-\frac{\pi}{3}}^{\textrm{B}}+\overleftrightarrow{\chi}_{-\frac{\pi}{6},-\frac{\pi}{3}}^{\textrm{B}}+\overleftrightarrow{\chi}_{\frac{\pi}{6},\frac{\pi}{2}}^{\textrm{B}}+\overleftrightarrow{\chi}_{\frac{\pi}{3},\frac{\pi}{2}}^{\textrm{B}} +\overleftrightarrow{\chi}_{\frac{\pi}{3},\frac{2\pi}{3}}^{\textrm{B}}\nonumber \\
 &+\overleftrightarrow{\chi}_{\frac{\pi}{2},\frac{\pi}{2}}^{\textrm{B}}+\overleftrightarrow{\chi}_{\frac{\pi}{2},\frac{2\pi}{3}}^{\textrm{B}}+\overleftrightarrow{\chi}_{\frac{\pi}{2},\frac{\text{5\ensuremath{\pi}}}{6}}^{\textrm{B}} +\frac{1}{2}\left(\overleftrightarrow{\chi}_{\frac{2\pi}{3},\frac{\pi}{2}}^{\textrm{B}}+\overleftrightarrow{\chi}_{\frac{2\pi}{3},\frac{2\pi}{3}}^{\textrm{B}}+\overleftrightarrow{\chi}_{\frac{2\pi}{3},\frac{5\pi}{6}}^{\textrm{B}}+\overleftrightarrow{\chi}_{\frac{2\pi}{3},\pi}^{\textrm{B}}\right).
\end{align}
In the first and third lines, the factors $1/2$ are due to half occupation of the levels at $\vert \pm\frac{2\pi}{3},\uparrow\rangle$. The components of the tensor
\begin{equation}
\overleftrightarrow{\chi}_{k,k'}^{\textrm{B}}=\sum_{i,j=x,y,z}\chi_{k,k'}^{\mathrm{B},i,j}\hat{e}_{i}\hat{e}_{j},
\end{equation}
are respectively
\begin{subequations}
\begin{align}
\text{\ensuremath{\chi}}_{k,k'}^{\mathrm{B},xx} & =\text{\ensuremath{\chi}}_{k,k'}^{\mathrm{B},yy}=-\alpha_{k,k'}^{2}\eta_{k,k'}^{\prime},\\
\text{\ensuremath{\chi}}_{k,k'}^{\mathrm{B},zz} & =-4\beta_{k,k'}^{2}\eta_{k,k'}^{\prime}(\delta_{k',k+\delta}+\delta_{k',k}),\\
\text{\ensuremath{\chi}}_{k,k'}^{\mathrm{B},xy} & =-\text{\ensuremath{\chi}}_{k,k'}^{\mathrm{B},xy} =i\alpha_{k,k'}^{2}\eta_{k,k'}^{\prime}(\delta{}_{k',k}-\delta{}_{k',k+\delta}-\delta_{k',k+2\delta}+\delta_{k',k-\delta}),\\
\text{\ensuremath{\chi}}_{k,k'}^{\mathrm{B},xz} & =-\chi_{k,k'}^{\mathrm{B},zx}=2i\alpha_{k,k'}\beta_{k,k'}\eta_{k,k'}^{\prime}(\delta{}_{k',k}-\delta{}_{k',k+\delta}),\\
\text{\ensuremath{\chi}}_{k,k'}^{\mathrm{B},yz} & =\text{\ensuremath{\chi}}_{k,k'}^{\mathrm{B},zy}=-2\alpha_{k,k'}\beta_{k,k'}\eta_{k,k'}^{\prime}(\delta_{k',k+\delta}+\delta_{k',k}).
\end{align}
\end{subequations}

For the transitions $\vert k,\downarrow\rangle\leftrightarrows\vert k',\uparrow\rangle$ ($k'=k,k+\delta$), the three eigenvalues are, respectively,
\begin{subequations}
\begin{eqnarray}
\mu_{r}^1&=&1-2\sum_{k\in\mathrm{HOMO}}\;\sum_{k'\in\mathrm{LUMO}}\alpha_{k,k'}^2\eta'_{k,k'},\\
\mu_{r}^2&=&1-2\sum_{k\in\mathrm{HOMO}}\;\sum_{k'\in\mathrm{LUMO}}\left(\beta_{k,k'}^2+\beta_{k,k'}\sqrt{2\alpha_{k,k'}^2+\beta_{k,k'}^2}\right)\eta'_{k,k'},\\
\mu_{r}^3&=&1-2\sum_{k\in\mathrm{HOMO}}\;\sum_{k'\in\mathrm{LUMO}}\left(\beta_{k,k'}^2-\beta_{k,k'}\sqrt{2\alpha_{k,k'}^2+\beta_{k,k'}^2}\right)\eta'_{k,k'}.
\end{eqnarray}
\end{subequations}
For the transitions $\vert k,\downarrow\rangle\leftrightarrows\vert k',\uparrow\rangle$ ($k'=k-\delta,k+2\delta$), the three eigenvalues are respectively
\begin{align}
\mu_{r}^1=1-2\sum_{k\in\mathrm{HOMO}}\;\sum_{k'\in\mathrm{LUMO}}\alpha_{k,k'}^2\eta'_{k,k'},\;\mu_{r}^2=\mu_{r}^3=1.
\end{align}
In both cases, the lowest eigenvalues is $\mu_{r}^1$,
because $(\beta_{k,k'}/\alpha_{k,k'})^2\sim\delta^2\ll1$. Therefore,
the bandwidth of the negative refraction is determined by
\begin{align}
\mu_{r}^1=1-2\sum_{k\in\mathrm{HOMO}}\;\sum_{k'\in\mathrm{LUMO}}\alpha_{k,k'}^2\eta'_{k,k'}=0.
\end{align}
Because the second terms of $\alpha_{k,k'}$ are proportional to $\delta$,
the maximum of $\alpha_{k,k}$ and $\alpha_{k,k+\delta}$ are generally larger than those of $\alpha_{k,k-\delta}$ and $\alpha_{k,k+2\delta}$. Assuming $\varepsilon=V$, we have
\begin{eqnarray}
\max\left(\alpha_{k,k}\right)&\simeq&\alpha_{\frac{\pi}{2}+\frac{\delta}{2},\frac{\pi}{2}+\frac{\delta}{2}}
=\frac{RV}{\hbar c}\left(1+2\sin\frac{3\delta}{4}\right)\simeq\frac{RV}{\hbar c}\left(1+\frac{6\delta}{4}\right)\simeq\frac{7RV}{4\hbar c},\\
\max\left(\alpha_{k,k+\delta}\right)&\simeq&\alpha_{-\frac{\pi}{2}+\frac{\delta}{2},-\frac{\pi}{2}+\frac{3\delta}{2}}
=\frac{RV}{\hbar c}\left(1+2\sin\frac{3\delta}{4}\right)\simeq\frac{RV}{\hbar c}\left(1+\frac{6\delta}{4}\right)\simeq\frac{7RV}{4\hbar c}.
\end{eqnarray}
In Ref.~\cite{Fang2016}, the bandwidth of the negative refraction is determined by
\begin{eqnarray}
\alpha_{0,0}=\frac{RV}{\hbar c}\left(1-2\sin\frac{\delta}{2}\sin\frac{3\delta}{4}\right)\simeq\frac{RV}{\hbar c}\left(1-2\times\frac{\delta}{2}\times\frac{3\delta}{4}\right)\simeq\frac{13RV}{16\hbar c}.
\end{eqnarray}
Since
\begin{eqnarray}
\left(\frac{\alpha_{\frac{\pi}{2}+\frac{\delta}{2},\frac{\pi}{2}+\frac{\delta}{2}}}{\alpha_{0,0}}\right)^2=
\left(\frac{\alpha_{-\frac{\pi}{2}+\frac{\delta}{2},-\frac{\pi}{2}+\frac{3\delta}{2}}}{\alpha_{0,0}}\right)^2\simeq5,
\end{eqnarray}
and there are 4 simultaneous transitions with the same frequency,
the bandwidth of the negative refraction is about 20 times that of Ref.~\cite{Fang2016}.

\section{Bandwidths of Negative Refraction for Different Transitions}

\label{app:Width}

As shown in Fig.~2 of the main text, there are 20 possible transitions on account of the initial condition, where there is an electron occupying the initial state, and the final condition, where there is no electron occupying the final state. By scanning over all transition frequencies, we obtain the bandwidths of the negative refraction for all transitions, as listed in Table~\ref{tab:FreWidth}.
Due to simultaneous multi-electron transitions,
the bandwidths of the negative refraction are generally larger than the corresponding
single-electron transition and that in Ref.~\cite{Fang2016}. Interestingly,
the bandwidths of five transitions are larger than the previous observation by one order of magnitude, as highlighted in red in Table~\ref{tab:FreWidth}.

\begin{table}[!ht]\centering
  \caption{The bandwidths of the negative refraction around different transition frequencies $\Delta_{if}$. In the third column, the participating transitions are explicitly given.  \label{tab:FreWidth}}
  \renewcommand\arraystretch{2}
  \begin{tabular}{ | r | c | l | }
    \hline
    $\;\;\;\Delta_{if}$ (eV)\;\;\; & \;\;\;bandwidth ($\mu$eV)\;\;\; & \;\;\; participating transitions\;\;\;\;\;\;\;\;\;\;\;\;\;\;\;\; \\ \hline
    1.736503\;\;\; & 9   &  \;\;\; $\vert\frac{2\pi}{3},\downarrow\rangle\rightleftarrows\vert\frac{\pi}{2},\uparrow\rangle$, $\vert-\frac{2\pi}{3},\downarrow\rangle\rightleftarrows\vert-\frac{\pi}{3},\uparrow\rangle$\\ \hline
    5.336503\;\;\; & {\color{red}80}  &  \;\;\; $\vert\frac{\pi}{2},\downarrow\rangle\rightleftarrows\vert\frac{\pi}{2},\uparrow\rangle$, $\vert-\frac{\pi}{2},\downarrow\rangle\rightleftarrows\vert-\frac{\pi}{3},\uparrow\rangle$\\ \hline
    5.463497\;\;\; & {\color{red}36}  & \;\;\; $\vert\frac{2\pi}{3},\downarrow\rangle\rightleftarrows\vert\frac{2\pi}{3},\uparrow\rangle$, $\vert-\frac{2\pi}{3},\downarrow\rangle\rightleftarrows\vert-\frac{\pi}{2},\uparrow\rangle$ \\ \hline
    8.691169\;\;\; & {\color{red}15}  & \;\;\; $\vert\frac{2\pi}{3},\downarrow\rangle\rightleftarrows\vert\frac{5\pi}{6},\uparrow\rangle$, $\vert-\frac{2\pi}{3},\downarrow\rangle\rightleftarrows\vert-\frac{2\pi}{3},\uparrow\rangle$ \\ \hline
    8.936503\;\;\; & {\color{red}39}  & \;\;\; $\vert\frac{\pi}{3},\downarrow\rangle\rightleftarrows\vert\frac{\pi}{2},\uparrow\rangle$, $\vert-\frac{\pi}{3},\downarrow\rangle\rightleftarrows\vert-\frac{\pi}{3},\uparrow\rangle$ \\ \hline
    9.063497\;\;\; & {\color{red}45}  & \;\;\; $\vert\frac{\pi}{2},\downarrow\rangle\rightleftarrows\vert\frac{2\pi}{3},\uparrow\rangle$, $\vert-\frac{\pi}{2},\downarrow\rangle\rightleftarrows\vert-\frac{\pi}{2},\uparrow\rangle$ \\ \hline
    10.554666\;\;\; & 3.7 & \;\;\; $\vert\frac{2\pi}{3},\downarrow\rangle\rightleftarrows\vert\pi,\uparrow\rangle$, $\vert-\frac{2\pi}{3},\downarrow\rangle\rightleftarrows\vert-\frac{5\pi}{6},\uparrow\rangle$ \\ \hline
    11.571886\;\;\; & 7   & \;\;\; $\vert\frac{\pi}{6},\downarrow\rangle\rightleftarrows\vert\frac{\pi}{2},\uparrow\rangle$, $\vert-\frac{\pi}{6},\downarrow\rangle\rightleftarrows\vert-\frac{\pi}{3},\uparrow\rangle$ \\ \hline
    12.291169\;\;\; & 6   & \;\;\; $\vert\frac{\pi}{2},\downarrow\rangle\rightleftarrows\vert\frac{5\pi}{6},\uparrow\rangle$, $\vert-\frac{\pi}{2},\downarrow\rangle\rightleftarrows\vert-\frac{2\pi}{3},\uparrow\rangle$ \\ \hline
    12.663497\;\;\; & 6   & \;\;\; $\vert\frac{\pi}{3},\downarrow\rangle\rightleftarrows\vert\frac{2\pi}{3},\uparrow\rangle$, $\vert-\frac{\pi}{3},\downarrow\rangle\rightleftarrows\vert-\frac{\pi}{2},\uparrow\rangle$ \\ \hline
  \end{tabular}
\end{table}

The comparison of the bandwidths of the negative refraction in the literatures and present paper is explicitly given in Table~\ref{tab:Literature}. The bandwidth has been broadened by at least \textit{two} orders of magnitude as compared to Refs.~\cite{Kastel2007-1,Kastel2007-2,Oktel2004,Thommen2006}.
\begin{table}[!ht]\centering
  \caption{Comparison of the bandwidths of the negative refraction in the literatures and present paper.  \label{tab:Literature}}
  \renewcommand\arraystretch{2}
  \begin{tabular}{ |c|c|c|c|c|c|}
    \hline
    Reference & Present & Ref.~\cite{Fang2016} & Refs.~\cite{Kastel2007-1,Kastel2007-2} & Ref.~\cite{Oktel2004} & Ref.~\cite{Thommen2006} \\ \hline
    bandwidth (Hz) & \;\;\; $1.93\times10^{10}$ \;\;\; & \;\;\; $0.966\times10^9$ \;\;\; & \;\;\; $\sim2\times10^5$ \;\;\; & \;\;\; $\sim5.03\times10^4$ \;\;\; & \;\;\; $1.6\times10^8$ \;\;\; \\ \hline
  \end{tabular}
\end{table}



\begin{thebibliography}{10}


\bibitem{Veselago1968}V. G. Veselago,
    The electrodynamics of substances with simultaneously negative values of $\varepsilon$ and $\mu$,
    \href{https://doi.org/10.1070/PU1968v010n04ABEH003699}
    {Sov. Phys. Uspekhi. \textbf{10}, 509 (1968)}.

\bibitem{Pendry2000}J. B. Pendry,
    Negative refraction makes a perfect lens,
    \href{https://doi.org/10.1103/PhysRevLett.85.3966}
    {Phys. Rev. Lett. \textbf{85}, 3966 (2000)}.

\bibitem{Smith2000}D. R. Smith, W. J. Padilla, D. C. Vier, S. C. Nemat-Nasser, and S. Schultz, Composite medium with simultaneously negative permeability and permittivity,
    \href{https://doi.org/10.1103/PhysRevLett.84.4184}
    {Phys. Rev. Lett. \textbf{84}, 4184 (2000)}.

\bibitem{Bliokh2008}K. Y. Bliokh, Y. P. Bliokh, V. Freilikher, S. Savel'ev, and F. Nori,
    Colloquium: Unusual resonators: Plasmonics, metamaterials, and random media,
    \href{https://doi.org/10.1103/RevModPhys.80.1201}
    {Rev. Mod. Phys. \textbf{80}, 1201 (2008)}.

\bibitem{Khorasaninejad2017}M. Khorasaninejad and F. Capasso,
    Metalenses: Versatile multifunctional photonic components,
    \href{https://doi.org/10.1126/science.aam8100}
    {Science \textbf{358}, 1146 (2017)}.

\bibitem{Minovich2015}A. E. Minovich, A. E. Miroshnichenko, A. Y. Bykov, T. V. Murzina, D. N. Neshev, and Y. S. Kivshar,
    Functional and nonlinear optical metasurfaces,
    \href{https://doi.org/10.1002/lpor.201400402}
    {Laser Photonics Rev. \textbf{9}, 195 (2015)}.

\bibitem{Zhao2016}R. K. Zhao, Y. Luo, and J. B. Pendry,
    Transformation optics applied to van der Waals interactions,
    \href{https://doi.org/10.1007/s11434-015-0958-x}
    {Sci. Bull. \textbf{61}, 59 (2016)}.

\bibitem{Shen2016}Y. Shen and Q. Ai,
    Optical properties of drug metabolites in latent fingermarks,
    \href{https://doi.org/10.1038/srep20336}
    {Sci. Rep. \textbf{6}, 20336 (2016)}.

\bibitem{Bliokh2013}Y. P. Bliokh, V. Freilikher, and F. Nori,
    Ballistic charge transport in graphene and light propagation in periodic dielectric structures with metamaterials: A comparative study,
    \href{https://doi.org/10.1103/PhysRevB.87.245134}
    {Phys. Rev. B \textbf{87}, 245134 (2013)}.

\bibitem{Kats2007}A. V. Kats, S. Savel'ev, V. A. Yampol'skii, and F. Nori,
    Left-handed interfaces for electromagnetic surface waves,
    \href{https://doi.org/10.1103/PhysRevLett.98.073901}
    {Phys. Rev. Lett. \textbf{98}, 073901 (2007)}.

\bibitem{Jiang2014}S.-C. Jiang, X. Xiong, Y.-S. Hu, Y.-H. Hu, G.-B. Ma,
    R.-W. Peng, C. Sun, and M. Wang,
    Controlling the polarization state of light with a dispersion-free metastructure,
    \href{https://doi.org/10.1103/PhysRevX.4.021026}
    {Phys. Rev. X \textbf{4}, 021026 (2014)}.

\bibitem{Fan2015}R.-H. Fan, Y. Zhou, X.-P. Ren, R.-W. Peng, S.-C. Jiang, D.-H. Xu,
    X. X. Xian, R. Huang, and M. Wang,
    Freely tunable broadband polarization rotator for terahertz waves,
    \href{https://doi.org/10.1002/adma.201404981}
    {Adv. Mater. \textbf{27}, 1201 (2015)}.

\bibitem{Leonhard2006}U. Leonhardt,
    Optical conformal mapping,
    \href{https://doi.org/10.1126/science.1126493}
    {Science \textbf{312}, 1777 (2006)}.

\bibitem{Pendry2006}J. B. Pendry, D. Schurig, and D. R. Smith,
    Controlling electromagnetic fields,
    \href{https://doi.org/10.1126/science.1125907}
    {Science \textbf{312}, 1780 (2006)}.


\bibitem{Xiong2013}X. Xiong, S.-C. Jiang, Y.-H. Hu, R.-W. Peng, and M. Wang,
    Structured metal film as a perfect absorber,
    \href{https://doi.org/10.1002/adma.201300223}
    {Adv. Mater. \textbf{25}, 3994 (2013)}.

\bibitem{Shen2014}Y. Shen, H. Y. Ko, Q. Ai, S. M. Peng, and B. Y. Jin,
    Molecular split-ring resonators based on metal string complexes,
    \href{https://doi.org/10.1021/jp410619d}
    {J. Phys. Chem. C \textbf{118}, 3766 (2014)}.

\bibitem{Pendry2004}J. B. Pendry,
    A chiral route to negative refraction,
    \href{https://doi.org/10.1126/science.1104467}
    {Science \textbf{306}, 1353 (2004)}.

\bibitem{Xiong2010}X. Xiong, W.-H. Sun, Y.-J. Bao, M. Wang, R.-W. Peng, C. Sun,
    X. Lu, J. Shao, Z.-F. Li, and N.-B. Ming,
    Construction of a chiral metamaterial with a U-shaped resonator assembly,
    \href{https://doi.org/10.1103/PhysRevB.81.075119}
    {Phys. Rev. B \textbf{81}, 075119 (2010)}.

\bibitem{Fisher1969}R. K. Fisher and R. W. Gould,
    Resonance cones in the field pattern of a short antenna in an anisotropic plasma,
    \href{https://doi.org/10.1103/PhysRevLett.22.1093}
    {Phys. Rev. Lett. \textbf{22}, 1093 (1969)}.

\bibitem{Smith2003}D. R. Smith and D. Schurig,
    Electromagnetic wave propagation in media with indefinite permittivity and permeability tensors,
    \href{https://doi.org/10.1103/PhysRevLett.90.077405}
    {Phys. Rev. Lett. \textbf{90}, 077405 (2003)}.

\bibitem{Rakhmanov2010}A. L. Rakhmanov, V. A. Yampol'skii, J. A. Fan, F. Capasso, and F. Nori,
    Layered superconductors as negative-refractive-index metamaterials,
    \href{https://doi.org/10.1103/PhysRevB.81.075101}
    {Phys. Rev. B \textbf{81}, 075101 (2010)}.

\bibitem{Ai2018}Q. Ai, P.-B. Li, W. Qin, C. P. Sun, and F. Nori,
    NV-Metamaterial: Tunable quantum hyperbolic metamaterial using nitrogen-vacancy
    centers in diamond,
    \href{https://arxiv.org/abs/1802.01280}
    {arXiv:1802.01280}.

\bibitem{Kastel2007-1}J. K\"{a}stel, M. Fleischhauer, S. F. Yelin, and R. L. Walsworth,
    Tunable negative refraction without absorption via electromagnetically induced chirality,
    \href{https://doi.org/10.1103/PhysRevLett.99.073602}
    {Phys. Rev. Lett. \textbf{99}, 073602 (2007)}.

\bibitem{Kastel2007-2} J. K\"{a}stel, M. Fleischhauer, and G. Juzeli\={u}nas,
    Local-field effects in magnetodielectric media: Negative refraction and absorption reduction,
    \href{https://doi.org/10.1103/PhysRevA.76.062509}
    {Phys. Rev. A \textbf{76}, 062509 (2007)}.

\bibitem{Qin2013}L. Qin, K. Zhang, R.-W. Peng, X. Xiong, W. Zhang,
    X.-R. Huang, and M. Wang,
    Optical-magnetism-induced transparency in a metamaterial,
    \href{https://doi.org/10.1103/PhysRevB.87.125136}
    {Phys. Rev. B \textbf{87}, 125136 (2013)}.

\bibitem{Wang2018}Y.-Y. Wang, J. Qiu, Y.-Q. Chu, M. Zhang, J.-M. Cai, Q. Ai, and F.-G. Deng,
    Dark state polarizing a nuclear spin in the vicinity of a nitrogen-vacancy center,
    \href{https://doi.org/10.1103/PhysRevA.97.042313}
    {Phys. Rev. A \textbf{97}, 042313 (2018)}.

\bibitem{Xiong2009}X. Xiong, W.-H. Sun, Y.-J. Bao, R.-W. Peng, M. Wang, C. Sun, X. Lu,
    J. Shao, Z.-F. Li, and N.-B. Ming,
    Switching the electric and magnetic responses in a metamaterial,
    \href{https://doi.org/10.1103/PhysRevB.80.201105}
    {Phys. Rev. B \textbf{80}, 201105(R) (2009)}.

\bibitem{Chang2010} C. W. Chang, M. Liu, S. Nam, S. Zhang, Y. Liu, G. Bartal, and X. Zhang,
    Optical M\"{o}bius symmetry in metamaterials,
    \href{https://doi.org/10.1103/PhysRevLett.105.235501}
    {Phys. Rev. Lett. \textbf{105}, 235501 (2010)}.

\bibitem{Krishnamoorthy2012} H. N. S. Krishnamoorthy, Z. Jacob, E. Narimanov, I. Kretzschmar, and V. M. Menon,
    Topological transitions in metamaterials,
    \href{https://doi.org/10.1126/science.1219171}
    {Science \textbf{336}, 205 (2012)}.

\bibitem{Fang2016}Y. N. Fang, Y. Shen, Q. Ai, and C. P. Sun,
    Negative refraction in M\"{o}bius molecules,
    \href{https://doi.org/10.1103/PhysRevA.94.043805}
    {Phys. Rev. A \textbf{94}, 043805 (2016)}.

\bibitem{Chen2005}Y. F. Chen, P. Fischer, and F. W. Wise,
    Negative refraction at optical frequencies in nonmagnetic two-component molecular media,
    \href{https://doi.org/10.1103/PhysRevLett.95.067402}
    {Phys. Rev. Lett. \textbf{95}, 067402 (2005)}.

\bibitem{Thommen2006}Q. Thommen and P. Mandel,
    Electromagnetically induced left handedness in optically excited four-level atomic media,
    \href{https://doi.org/10.1103/PhysRevLett.96.053601}
    {Phys. Rev. Lett. \textbf{96}, 053601 (2006)}.

\bibitem{Orth2013}P. P. Orth, R. Hennig, C. H. Keitel, and J. Evers,
    Negative refraction with tunable absorption in an active dense gas of atoms,
    \href{https://doi.org/10.1088/1367-2630/15/1/013027}
    {New J. Phys. \textbf{15}, 013027 (2013)}.

\bibitem{Oktel2004} M. \"{O}. Oktel and \"{O}. E. M\"{u}stecapl{\i}o\u{g}lu,
    Electromagnetically induced left-handedness in a dense gas of three-level atoms,
    \href{https://doi.org/10.1103/PhysRevA.70.053806}
    {Phys. Rev. A \textbf{70}, 053806 (2004)}.

\bibitem{Ceulemans1998}A. Ceulemans, L. F. Chibotaru, and P. W. Fowler,
    Molecular anapole moments,
    \href{https://doi.org/10.1103/PhysRevLett.80.1861}
    {Phys. Rev. Lett. \textbf{80}, 1861 (1998)}.

\bibitem{Zagoskin2015}A. M. Zagoskin, A. Chipouline, E. Il'ichev, J. R. Johansson,
    and F. Nori,
    Toroidal qubits: Naturally-decoupled quiet artificial atoms,
    \href{https://doi.org/10.1038/srep16934}
    {Sci. Rep. \textbf{5}, 16934 (2015)}.

\bibitem{Heilbronner1964}E. Heilbronner,
    H\"{u}ckel molecular orbitals of M\"{o}bius-type conformations of annulenes,
    \href{https://doi.org/10.1016/S0040-4039(01)89474-0}
    {Tetrahedron Lett. \textbf{5}, 1923 (1964)}.

\bibitem{Walba1993}D. M. Walba, T. C. Homan, R. M. Richards, and R. C. Haltiwanger,
    Topological stereochemistry. IX: Synthesis and cutting in half of a molecular M\"{o}bius strip, \href{https://dx.doi.org/10.1039/C7NJ02828H}
    {New J. Chem. \textbf{17}, 661 (1993)}.

\bibitem{Ajami2003}D. Ajami, O. Oeckler, A. Simon, and R. Herges,
    Synthesis of a M\"{o}bius aromatic hydrocarbon,
    \href{https://doi.org/10.1038/nature02224}
    {Nature \textbf{426}, 819 (2003)}.

\bibitem{Yoneda2014}T. Yoneda, Y. M. Sung, J. M. Lim, D. Kim, and A. Osuka,
    Pd$^{\mathrm{II}}$ complexes of {[}44{]}- and {[}46{]} decaphyrins:
    The largest H\"{u}ckel aromatic and antiaromatic, and M\"{o}bius aromatic macrocycles,
    \href{https://doi.org/10.1002/anie.201408506}
    {Angew. Chem. Int. Ed. \textbf{126}, 13385 (2014)}.

\bibitem{Poddar2014}A. K. Poddar and U. L. Rohde,
    M\"{o}bius strips and metamaterial symmetry: Theory and applications,
    \href{https://doi.org/10.1109/FCS.2014.6859924}
    {Microwave J. \textbf{57}, 76 (2014)}.

\bibitem{Balzani2008}V. Balzani, A. Credi, and M. Venturi,
    \textit{Molecular Devices and Machines. Concepts and Perspectives for the Nanoworld}
    (VCH-Wiley, Weinheim, 2008).

\bibitem{Yamashiroa2004}A. Yamashiroa, Y. Shimoia, K. Harigayaa, and K. Wakabayashi,
    Novel electronic states in graphene ribbons--competing spin and charge orders,
    \href{https://doi.org/10.1016/j.physe.2003.12.100}
    {Physica E \textbf{22}, 688 (2004)}.

\bibitem{Zhao2009}N. Zhao, H. Dong, S. Yang, and C. P. Sun,
    Observable topological effects in molecular devices with M\"{o}bius topology,
    \href{https://doi.org/10.1103/PhysRevB.79.125440}
    {Phys. Rev. B \textbf{79}, 125440 (2009)}.

\bibitem{Pond2000}J. M. Pond,
    M\"{o}bius dual-mode resonators and bandpass filters,
    \href{https://doi.org/10.1109/22.898999}
    {IEEE Trans. Microw. Theory Tech. \textbf{48}, 2465 (2000)}.

\bibitem{Guo2009}Z. L. Guo, Z. R. Gong, H. Dong, and C. P. Sun,
    M\"{o}bius graphene strip as a topological insulator,
    \href{https://doi.org/10.1103/PhysRevB.80.195310}
    {Phys. Rev. B \textbf{80}, 195310 (2009)}.

\bibitem{Lukin2005}O. Lukin and F. V\"{o}gtle,
    Knotting and threading of molecules: Chemistry and chirality of molecular knots and their assemblies,
    \href{https://doi.org/10.1002/anie.200460312}
    {Angew. Chem. Int. Ed. \textbf{44}, 1456 (2005)}.

\bibitem{Xu2018}L. Xu, Z. R. Gong, M. J. Tao, and Q. Ai,
    Artificial light harvesting by dimerized M\"{o}bius ring,
    \href{https://doi.org/10.1103/PhysRevE.97.042124}
    {Phys. Rev. E \textbf{97}, 042124 (2018)}.

\bibitem{Lambert2013}N. Lambert, Y.-N. Chen, Y.-C. Cheng, G.-Y. Chen, and F. Nori,
    Quantum biology,
    \href{https://doi.org/10.1038/nphys2474}
    {\textit{Nature Phys.} \textbf{9,} 10 (2013)}.

\bibitem{Landau1995}L. D. Landau, E. M. Lifshitz, and L. P. Pitaevskii,
    \textit{Electrodynamics of Continuous Media} 2nd Ed.
    (Butterworth Heinmann, Oxford, 1995).

\bibitem{Jackson1999}J. D. Jackson,
    \textit{Classical Electrodynamics} 3rd Ed.
    (John Wiley, United States, 1999).

\bibitem{Kubo1985}R. Kubo, M. Toda, and N. Hashitsume,
    \textit{Statistical Physics II Nonequlibirum Statistical Mechanics}
    (Springer-Verlag, Berlin Heidelberg, 1985).

\bibitem{SuppMat}See Supplemental material for details of calculation,
    which includes Refs.~\cite{Fang2016,Salem1972,Landau1995,Jackson1999,Kubo1985,Ai2010}.

\bibitem{Jang2018}S. J. Jang and B. Mennucci,
    Delocalized excitons in natural light harvesting complexes,
    \href{https://www.arxiv.org/abs/1804.09711}
    {to be published in Rev. Mod. Phys., also available at arXiv:1804.09711}.

\bibitem{Novoderezhkin2010}V. I. Novoderezhkin and R. van Grondelle,
    Physical origins and models of energy transfer in photosynthetic light-harvesting,
    \href{https://doi.org/10.1039/C003025B}
    {Phys. Chem. Chem. Phys. \textbf{12}, 7352 (2010)}.

\bibitem{Cheng2006}Y. C. Cheng and R. J. Silbey,
    Coherence in the B800 ring of purple bacteria LH2,
    \href{https://doi.org/10.1103/PhysRevLett.96.028103}
    {Phys. Rev. Lett. \textbf{96}, 028103 (2006)}.

\bibitem{Salem1972}L. Salem,
    \textit{The Molecular Orbital Theory of Conjugated Systems}
    (Benjamin, Reading, MA, 1972).

\bibitem{Ai2010}Q. Ai, Y. Li, H. Zheng, and C. P. Sun,
    Quantum anti-Zeno effect without rotating wave approximation,
    \href{https://doi.org/10.1103/PhysRevA.81.042116}
    {Phys. Rev. A \textbf{81}, 042116 (2010)}.

\bibitem{Wang2017}X. Wang, A. Miranowicz, H. R. Li, F. Nori,
    Observing pure effects of counter-rotating terms without ultrastrong coupling: A single photon can simultaneously excite two qubits,
    \href{https://doi.org/10.1103/PhysRevA.96.063820}
    {Phys. Rev. A \textbf{96}, 063820 (2017)}.

\bibitem{Greenwood1972}H. H. Greenwood,
    \textit{Computing Methods in Quantum Organic Chemistry}
    (Wiley-Interscience, Germany, 1972).

%
%
%
%
%
%
%
%
%
%





\end{thebibliography}

\begin{thebibliography}{10}


\bibitem{Kubo1985}R. Kubo, M. Toda, and N. Hashitsume,
    \textit{Statistical Physics II Nonequlibirum Statistical Mechanics}
    (Springer-Verlag, Berlin Heidelberg, 1985).

\bibitem{Ai2010}Q. Ai, Y. Li, H. Zheng, and C. P. Sun,
    Quantum anti-Zeno effect without rotating wave approximation,
    \href{https://doi.org/10.1103/PhysRevA.81.042116}
    {Phys. Rev. A \textbf{81}, 042116 (2010)}.

\bibitem{Jackson1999}J. D. Jackson,
    \textit{Classical Electrodynamics} 3rd Ed.
    (John Wiley, United States, 1999).

\bibitem{Landau1995}L. D. Landau, E. M. Lifshitz, and L. P. Pitaevskii,
    \textit{Electrodynamics of Continuous Media} 2nd Ed.
    (Butterworth Heinmann, Oxford, 1995).

\bibitem{Salem1972}L. Salem,
    \textit{The Molecular Orbital Theory of Conjugated Systems}
    (Benjamin, Reading, MA, 1972).

\bibitem{Fang2016}Y. N. Fang, Y. Shen, Q. Ai, and C. P. Sun,
    Negative refraction in M\"{o}bius molecules,
    \href{https://doi.org/10.1103/PhysRevA.94.043805}
    {Phys. Rev. A \textbf{94}, 043805 (2016)}.

\bibitem{Kastel2007-1}J. K\"{a}stel, M. Fleischhauer, S. F. Yelin, and R. L. Walsworth,
    Tunable negative refraction without absorption via electromagnetically induced chirality,
    \href{https://doi.org/10.1103/PhysRevLett.99.073602}
    {Phys. Rev. Lett. \textbf{99}, 073602 (2007)}.

\bibitem{Kastel2007-2}J. K\"{a}stel, M. Fleischhauer, and G. Juzeli\={u}nas,
    Local-field effects in magnetodielectric media: Negative refraction and absorption reduction,
    \href{https://doi.org/10.1103/PhysRevA.76.062509}
    {Phys. Rev. A \textbf{76}, 062509 (2007)}.

\bibitem{Oktel2004}M. \"{O}. Oktel and \"{O}. E. M\"{u}stecapl{\i}o\u{g}lu,
    Electromagnetically induced left-handedness in a dense gas of three-level atoms,
    \href{https://doi.org/10.1103/PhysRevA.70.053806}
    {Phys. Rev. A \textbf{70}, 053806 (2004)}.

\bibitem{Thommen2006}Q. Thommen and P. Mandel,
    Electromagnetically induced left handedness in optically excited four-level atomic media,
    \href{https://doi.org/10.1103/PhysRevLett.96.053601}
    {Phys. Rev. Lett. \textbf{96}, 053601 (2006)}.


\end{thebibliography}
\end{document}